\documentclass[pdflatex,iicol,sn-mathphys-num]{sn-jnl}
\usepackage[T1]{fontenc}
\usepackage{amsmath,amssymb,mathtools,bm}
\usepackage{microtype}
\usepackage{booktabs,graphicx}
\usepackage{balance}
\hypersetup{pdftitle={Coupled membrane response and horizon dissipation in Einstein-scalar-Gauss-Bonnet gravity},pdfauthor={Yan-Heng Jin and Wen-Biao Han}}
\AtBeginDocument{\setlength{\mathindent}{0pt}}
\newcommand{\GB}{\mathcal G}
\newcommand{\rh}{r_H}

\begin{document}
\title[Coupled membrane response and horizon dissipation in EsGB gravity]{Coupled membrane response and horizon dissipation in four-dimensional Einstein--scalar--Gauss--Bonnet gravity}
\author[1,2,3]{\fnm{Yan-Heng} \sur{Jin}}
\author*[1,4,5]{\fnm{Wen-Biao} \sur{Han}}\email{wbhan@shao.ac.cn}
\affil[1]{\orgdiv{School of Fundamental Physics and Mathematical Sciences}, \orgname{Hangzhou Institute for Advanced Study, University of Chinese Academy of Sciences}, \orgaddress{\city{Hangzhou} \postcode{310024}, \country{China}}}
\affil[2]{\orgname{Institute of Theoretical Physics, Chinese Academy of Sciences}, \orgaddress{\city{Beijing} \postcode{100190}, \country{China}}}
\affil[3]{\orgname{University of Chinese Academy of Sciences}, \orgaddress{\city{Beijing} \postcode{100049}, \country{China}}}
\affil[4]{\orgdiv{State Key Laboratory of Radio Astronomy and Technology}, \orgname{Shanghai Astronomical Observatory, Chinese Academy of Sciences}, \orgaddress{\street{80 Nandan Road}, \city{Shanghai} \postcode{200030}, \country{China}}}
\affil[5]{\orgdiv{School of Astronomy and Space Science}, \orgname{University of Chinese Academy of Sciences}, \orgaddress{\city{Beijing} \postcode{100049}, \country{China}}}
\abstract{
We develop an action-based membrane paradigm for nonextremal, static, spherically symmetric black holes in four-dimensional Einstein–scalar–Gauss–Bonnet (EsGB) gravity, describing their interaction with external perturbations through an effective fluid membrane. We calculate the linear membrane responses to perturbations and use them to analyze the work done on the membrane and the energy absorbed by the horizon. Horizon energy absorption receives contributions from geometric shear and scalar interactions. The scalar contribution includes both direct work on the membrane and work associated with the induced horizon expansion. We then establish a balance relation between the integrated horizon absorption and the second-order change in Wald entropy between stationary horizon cuts. Applying this framework to slow dynamics near a simple turning point, we obtain the leading exponential growth or decay rate of spherically symmetric scalar perturbations without fitting a dynamical damping coefficient.
}
\maketitle

\section{Introduction}
\label{sec:introduction}

The nonlinear field equations of a gravitational theory govern both its
black-hole equilibria and their response to disturbances.
Four-dimensional Einstein--scalar--Gauss--Bonnet
(EsGB) gravity provides a concrete example. A nonminimal coupling
$f(\phi)\mathcal G$ between a dynamical scalar $\phi$ and the
Gauss--Bonnet invariant $\mathcal G$, with coupling function $f$, introduces
scalar--curvature interactions into the field equations. For suitable
coupling functions, the theory admits branches of black holes with scalar
hair and spontaneous scalarization
\cite{DonevaYazadjiev2018,Silva2018}. Nonlinear evolutions further show
that the coupling affects the growth and saturation of scalar instabilities
and the resulting black-hole state \cite{EastRipley2021}.

Equilibrium solutions alone do not determine how a disturbed black hole
exchanges energy or how quickly it evolves. In general relativity,
linear quasinormal-mode models can describe early ringdown despite
nonlinear changes to the mode amplitudes \cite{Giesler2019,Zhu2024},
and inverse energy transfer has been identified in several black-hole
and gravitational-wave settings \cite{Yang2015,Ma2026,Ianniccari2025}.
These studies motivate examining how interactions between the geometry
and other fields, together with the channels for energy loss, govern
the response and evolution of a perturbed black hole.

The membrane paradigm expresses horizon dynamics through effective surface
responses and their work. For an exterior observer, sources on a timelike
stretched horizon represent the influence of the excised interior region.
In Einstein gravity,
the horizon limit of the membrane stress tensor admits a viscous-fluid
interpretation, with energy and momentum balances expressed in terms of
pressure, shear, and energy exchange
\cite{PriceThorne1986,ParikhWilczek1998}. The action-based formulation
identifies the sources from the boundary variation of the exterior problem
\cite{ParikhWilczek1998}. This description lets us study the work of
the geometric and scalar responses together, and compare their balance
structure and response coefficients with the Einstein-gravity limit.
Earlier studies of Einstein--Gauss--Bonnet membranes
\cite{JMS2017,FGS2018} and scalar--Gauss--Bonnet transport in
anti-de Sitter backgrounds \cite{Tong2025} illustrate how curvature
couplings enter fluid descriptions of gravitational systems.

Our aim is to determine how scalar--curvature coupling changes the geometric
and scalar responses of EsGB horizons, how their combined work determines
absorption and entropy balance, and how that absorption enters slow growth
or decay near a simple turning point. The EsGB boundary variation supplies
two conjugate membrane quantities: a stress tensor associated with changes
of the induced geometry and a scalar response associated with changes of the
boundary scalar. We construct them using the known Dirichlet boundary terms
and canonical momenta \cite{JulieBerti2020}. Their linear responses
determine the leading quadratic work when contracted with the corresponding
first-order geometric and scalar rates.

We evaluate the linear responses about regular, nonextremal, asymptotically
flat, static, spherically symmetric black holes, allowing angular dependence in the
perturbations. The membrane energy and momentum balances agree with the
corresponding null and null-spatial projections of the linearized metric
field equations on the horizon. After separating the equilibrium work and
imposing the teleological boundary condition of future stationarity, the
coupled work yields a positive quadratic absorbed power with distinct shear and scalar weights,
together with a total time derivative describing temporal storage. The
scalar absorption weight includes the work of the induced expansion. Using
an intrinsic extension of stationary Wald entropy
\cite{Wald1993,JulieBerti2019,JKM1994}, we relate this power to the
quadratic entropy balance between horizon cuts. For stationary initial and
final cuts, the integrated absorbed power equals the horizon temperature
times the second-order Wald-entropy change; for general cuts, the balance
also retains entropy-rate and temporal-storage endpoint terms.

Building on radial stability analyses
\cite{BlazquezSalcedo2018} and effective descriptions of scalarization
\cite{Khalil2019,Khalil2022}, we apply the scalar absorption to a slow
spherical mode near a simple turning point of a static family at fixed
Wald entropy. This family determines an effective potential whose negative
derivative gives the effective force. Combining this force with horizon
absorption and outgoing scalar radiation gives the leading exponential
growth or decay rate.

Relative to the constant-coupling Einstein--Gauss--Bonnet membrane
\cite{JMS2017} and Lovelock entropy-production analysis \cite{FGS2018},
we resolve the coupled geometric and scalar work and its contribution to
horizon absorption. Hollands, Kov\'acs and Reall construct horizon entropy
in an effective field theory (EFT) expansion \cite{HKR2022}; our calculation
retains the full coupling dependence within the specified EsGB action.
The detailed entropy comparison is given in Sec.~\ref{sec:entropy}.

Section~\ref{sec:model} constructs the membrane description, evaluates
the linear horizon response, and checks the energy and momentum balances
against the corresponding projections of the linearized metric field
equations. Section~\ref{sec:entropy} derives the absorption and quadratic
entropy balance from the coupled work. Section~\ref{sec:slow} applies
the scalar absorption to slow spherical dynamics near a simple turning
point. Section~\ref{sec:discussion} discusses the results and their scope;
Appendices~\ref{app:boundary} and~\ref{app:entropy} contain the boundary
and entropy derivations.

\section{Membrane paradigm for EsGB black holes}
\label{sec:model}

In this section, we construct an action-based membrane description for four-dimensional EsGB black holes. We evaluate the linear perturbations of the membrane sources about regular, nonextremal, static, spherically symmetric backgrounds, allowing for angular dependence. We then derive the membrane energy and momentum balances and verify their agreement with the corresponding horizon projections of the linearized metric field equations.

\subsection{Exterior description in EsGB gravity}
\label{sec:exterior-description}

The membrane paradigm treats a black hole, from the perspective of an exterior observer, as an effective fluid residing on a timelike surface $\Sigma$ just outside the true event horizon, known as the stretched horizon~\cite{PriceThorne1986,ParikhWilczek1998}. Effective sources on this surface represent the influence of the omitted interior region on the exterior fields.

Following the action-based formulation of Parikh and Wilczek~\cite{ParikhWilczek1998}, restricting the bulk action to the exterior of $\Sigma$ generates an inner-boundary variation. A membrane contribution cancels the remaining boundary variation. The coefficients of the boundary metric and scalar variations define the effective surface sources: a stress tensor conjugate to the induced boundary metric, and a scalar response conjugate to the boundary scalar field.

We consider four-dimensional Einstein-scalar-Gauss-Bonnet gravity with the bulk action
\begin{equation}
 \begin{aligned}
 S_{\rm bulk}&=\frac{1}{16\pi G}\int_M \mathrm d^4x\sqrt{-g}\,
 \Bigl[R-2\nabla_\mu\phi\nabla^\mu\phi\\
 &\hspace{35mm}+\alpha f(\phi)\GB\Bigr],\\
 \GB&\equiv R_{\mu\nu\rho\sigma}R^{\mu\nu\rho\sigma}
          -4R_{\mu\nu}R^{\mu\nu}+R^2.
 \end{aligned}
 \label{eq:m-action}
\end{equation}

Here $M$ denotes the spacetime integration region, $G$ is Newton's constant, $g=\det g_{\mu\nu}$, and $\nabla_\mu$ is the Levi-Civita connection of the spacetime metric. The scalar $\phi$ and coupling function $f(\phi)$ are dimensionless, whereas the coupling constant $\alpha$ has dimensions of length squared. Since the bare Gauss-Bonnet term $\GB$ is topological in four dimensions, the scalar dependence in $f(\phi)$ is required to induce local scalar-curvature interactions.

We use the metric signature $(-,+,+,+)$ and the Riemann convention $R^\mu{}_{\nu\rho\sigma} = \partial_\rho\Gamma^\mu_{\sigma\nu} - \partial_\sigma\Gamma^\mu_{\rho\nu} + \Gamma^\mu_{\rho\lambda}\Gamma^\lambda_{\sigma\nu} - \Gamma^\mu_{\sigma\lambda}\Gamma^\lambda_{\rho\nu}$. Greek indices refer to spacetime, $a,b,\ldots$ to the three-dimensional membrane, and $A,B,\ldots$ to its two-dimensional spatial cuts. The notation $f_{,\phi}$ denotes differentiation of the coupling function with respect to the scalar field; primes denote derivatives with respect to the static radial coordinate $r$.

Varying the action yields the equations of motion
\begin{align}
 E_{\mu\nu}={}&G_{\mu\nu}-2\nabla_\mu\phi\nabla_\nu\phi
       +g_{\mu\nu}(\nabla\phi)^2\nonumber\\
 &+4\alpha P_{\mu\rho\nu\sigma}\nabla^\rho\nabla^\sigma f=0,
 \label{eq:m-metric}\\
 E_\phi={}&4\Box\phi+\alpha f_{,\phi}\GB=0,
 \label{eq:m-field}
\end{align}

where $G_{\mu\nu}=R_{\mu\nu}-\tfrac12g_{\mu\nu}R$ is the Einstein tensor, $R_{\mu\nu}$ and $R$ are the Ricci tensor and scalar, $\Box = \nabla_\mu\nabla^\mu$, and
\begin{equation}
 \begin{aligned}
 P_{\mu\nu\rho\sigma}\equiv{}&R_{\mu\nu\rho\sigma}
       +g_{\mu\sigma}R_{\nu\rho}-g_{\mu\rho}R_{\nu\sigma}\\
 &-g_{\nu\sigma}R_{\mu\rho}+g_{\nu\rho}R_{\mu\sigma}\\
 &+\tfrac12 R(g_{\mu\rho}g_{\nu\sigma}-g_{\mu\sigma}g_{\nu\rho}).
 \end{aligned}
 \label{eq:m-curvature}
\end{equation}

The tensor $P_{\mu\nu\rho\sigma}$ inherits the Riemann symmetries and is divergence-free ($\nabla^\mu P_{\mu\nu\rho\sigma} = 0$). These equations determine the background horizon data and provide the bulk check for the membrane balances in Sec.~\ref{sec:finite-balances}.

\subsection{Boundary variation and membrane sources}
\label{sec:membrane-sources}

The Dirichlet action first eliminates normal derivatives of the metric variation, leaving only terms conjugate to the induced boundary fields. The membrane action then supplies the required response when these boundary fields are allowed to vary.

Let $n^\mu$ be the unit outward normal pointing from the omitted interior region to the exterior. We define the induced metric and extrinsic curvature as
\begin{equation}
 \begin{aligned}
 h_{\mu\nu}&\equiv g_{\mu\nu}-n_\mu n_\nu,\\
 K_{ab}&\equiv h_a{}^\mu h_b{}^\nu\nabla_\mu n_\nu,
 \qquad K\equiv h^{ab}K_{ab}.
 \end{aligned}
 \label{eq:m-boundary-geometry}
\end{equation}

Let $D_a$ be the connection compatible with $h_{ab}$, and $\widehat G_{ab} = \widehat R_{ab} - \frac{1}{2}h_{ab}\widehat R$ be the intrinsic boundary Einstein tensor. With the outward normal of each region used for its respective boundary integral, the EsGB Dirichlet action is~\cite{JulieBerti2020}
\begin{equation}
 \begin{aligned}
 S_D={}&S_{\rm bulk}+\frac{1}{8\pi G}\int_{\partial M}\mathrm d^3x\sqrt{-h}\\
 &\quad\times\left[K+2\alpha f(\phi)(J-2\widehat G^{ab}K_{ab})\right],\\
 J\equiv{}&\tfrac13\left(3KK_{ab}K^{ab}
       -2K_a{}^bK_b{}^cK_c{}^a-K^3\right).
 \end{aligned}
 \label{eq:m-dirichlet}
\end{equation}

This boundary completion ensures a well-posed Dirichlet problem when $h_{ab}$ and $\phi$ are fixed. Variations at the other boundaries and at joints are held fixed. For the exterior membrane problem, however, we retain the metric and scalar variations on $\Sigma$.

Smoothly matching the interior and exterior solutions demands the cancellation of their common boundary variations. The membrane effectively replaces the omitted interior region to balance the exterior Dirichlet variation:
\begin{equation}
	S_{\rm eff} = S_{D,\rm out} + S_{\rm mem},\qquad
	\delta S_{\rm mem}\big|_\Sigma = -\delta S_{D,\rm out}\big|_\Sigma.
	\label{eq:m-matching}
\end{equation}
Evaluating the omitted interior Dirichlet action on a solution regular at the future horizon realizes this variation. Its boundary momenta define the membrane stress tensor $t_{ab}$ and scalar response $j_\phi$:
\begin{equation}
 \begin{aligned}
 \delta S_{\rm mem}\big|_\Sigma
 &=\delta S_{D,\rm in}^{\rm on\mbox{-}shell}\big|_\Sigma\\
 &=\int_\Sigma \mathrm d^3x\sqrt{-h}
       \left(-\tfrac12t_{ab}\delta h^{ab}+j_\phi\delta\phi\right).
 \end{aligned}
 \label{eq:m-variation}
\end{equation}

Using the canonical momenta derived in Ref.~\cite{JulieBerti2020} under our normal convention, we obtain
\begin{strip}
\begin{align}
	8\pi G\,t_{ab} &= K h_{ab} - K_{ab} - 4\alpha\Bigl\{
	h_{ab}\bigl[K D^2f - K^{cd}D_cD_df\bigr] - K_{ab}D^2f - K D_aD_bf \nonumber\\
	&\quad + 2K_{c(a}D_{b)}D^cf + (n^\mu\nabla_\mu f) \Bigl[\widehat G_{ab} + K_a{}^cK_{cb} - K K_{ab} + \frac{1}{2}(K^2-K_{cd}K^{cd})h_{ab}\Bigr]\Bigr\}, \label{eq:m-stress} \\
	4\pi G\,j_\phi &= -n^\mu\nabla_\mu\phi + \alpha f_{,\phi}\left(J - 2\widehat G^{ab}K_{ab}\right), \label{eq:m-scalar}
\end{align}
\end{strip}
where $D^2f \equiv h^{ab}D_aD_bf$, and parentheses denote index symmetrization. In $t_{ab}$, the Hessian terms describe the variation of the coupling along $\Sigma$, while $n^\mu\nabla_\mu f$ measures its transverse gradient. Appendix~\ref{app:boundary} provides the explicit contractions and direct scalar variation. In four dimensions, the topological nature of the Gauss-Bonnet term ensures that the metric response depends exclusively on derivatives of $f$; the undifferentiated $f$ term vanishes identically due to antisymmetric boundary properties. Setting $\alpha=0$ recovers the Einstein-gravity responses with a minimally coupled scalar:
\begin{equation}
	t_{ab} = \frac{K h_{ab}-K_{ab}}{8\pi G},\qquad
	j_\phi = -\frac{n^\mu\nabla_\mu\phi}{4\pi G}.
	\label{eq:m-einstein-response}
\end{equation}

To establish a fluid interpretation, we introduce membrane observers with a unit timelike velocity $u^a$ orthogonal to the spatial sections of $\Sigma$. With the spatial projector $\gamma_{ab} \equiv h_{ab} + u_au_b$, the stress tensor decomposes as
\begin{equation}
	\begin{split}
		t_{ab} &= \rho u_au_b + p\gamma_{ab} + 2u_{(a}\mathcal Q_{b)} + \Pi_{ab}, \\
		\rho &\equiv t_{ab}u^au^b, \qquad \mathcal Q_a \equiv -\gamma_a{}^ct_{cd}u^d, \\
		p &\equiv \frac{1}{2}\gamma^{ab}t_{ab}, \qquad u^a\Pi_{ab} = \gamma^{ab}\Pi_{ab} = 0.
	\end{split}
	\label{eq:m-observer-stress}
\end{equation}
These quantities define the surface energy density $\rho$, spatial energy flux $\mathcal Q_a$, isotropic stress $p$ (scaled by $1/2$ for two spatial dimensions), and traceless spatial stress $\Pi_{ab}$. The conjugate response $j_\phi$ enters the scalar work and force in the surface balance equations derived below.

\subsection{Static spherical backgrounds and the horizon limit}
\label{sec:horizon-limit}

We evaluate the linear membrane response about a regular, nonextremal, asymptotically flat, static, spherically symmetric black hole, retaining angular dependence in the perturbations. The background fields are
\begin{equation}
 \begin{aligned}
 ds_0^2&=-e^{2\delta(r)}N(r)dt^2+\frac{dr^2}{N(r)}+r^2d\Omega_2^2,\\
 \phi&=\phi(r).
 \end{aligned}
 \label{eq:m-background}
\end{equation}

Here $r$ is the areal radius, so each symmetry sphere has area $4\pi r^2$, and $d\Omega_2^2=\Omega_{AB}dx^Adx^B$ is the line element of the unit two-sphere with metric $\Omega_{AB}$. At infinity we impose $N\to1$, $\delta\to0$, and $\phi\to\phi_\infty$, where $\phi_\infty$ is the prescribed asymptotic scalar value. The choice $\delta(\infty)=0$ fixes the background time normalization. At the event horizon $r=\rh$, the metric function $N(r)$ exhibits a simple zero with $N'_H>0$, while $\delta(r)$, $\phi(r)$, and their radial derivatives remain finite. The bulk field equations impose regularity conditions at the horizon~\cite{JulieBerti2019,JSBY2022}:
\begin{align}
	(\rh+2\alpha f_{,\phi,H}\phi'_H)N'_H &= 1, & \phi'_H &= -\frac{3\alpha f_{,\phi,H}N'_H}{\rh^2}.
	\label{eq:m-regularity}
\end{align}
Focusing on the physical branch that continuously connects to the general relativity limit ($\alpha \to 0$), the required horizon data (detailed in Appendix~\ref{app:horizon-data}) evaluate to 
\begin{align}
 \Delta_H&=\sqrt{1-\frac{24\alpha^2f_{,\phi,H}^{\,2}}{\rh^4}}>0,
 \nonumber\\
 \phi'_H&=-\frac{6\alpha f_{,\phi,H}}{\rh^3(1+\Delta_H)},\nonumber\\
 N'_H&=\frac1{\rh+2\alpha f_{,\phi,H}\phi'_H},
 \quad\kappa=\tfrac12e^{\delta_H}N'_H,\nonumber\\
 \mathcal A&=1-2\alpha N'_Hf_{,\phi,H}\phi'_H
       =\rh N'_H=\frac2{1+\Delta_H}.
 \label{eq:m-horizon}
\end{align}

Here, the subscript $H$ denotes quantities evaluated at the horizon, and $\kappa$ is the background surface gravity. The regularity parameters $\Delta_H$ and $\mathcal A$ approach unity in the vanishing-coupling limit. We restrict attention to the nondegenerate branch with $\Delta_H>0$ and assume that the background admits a regular asymptotically flat exterior.

Since spacetime perturbations deform the event horizon, we adopt Gaussian null coordinates $(v, \varrho, x^A)$ adapted to the perturbed horizon generators $\ell=\partial_v$ (located at $\varrho=0$):
\begin{equation}
 \begin{aligned}
 ds^2={}&2dv\,d\varrho-2\kappa\varrho\,dv^2
          +2\varrho\beta_A\,dv\,dx^A\\
 &+q_{AB}(v,\varrho,x)dx^Adx^B+O(\varrho^2).
 \end{aligned}
 \label{eq:h-coordinates}
\end{equation}

In this line element, $\beta_A$ serves as the transverse shift vector, and on the null hypersurface $\varrho=0$, $q_{AB}$ reduces to the induced metric of a spatial horizon cut. We fix the nonaffinity of the generators to match the background surface gravity $\kappa$; this gauge choice is permitted locally because an infinitesimal generator rescaling $\ell^\mu\mapsto(1+\gamma)\ell^\mu$, with normalization perturbation $\gamma(v,x^A)$, shifts the nonaffinity by $(\partial_v+\kappa)\gamma$ at linear order. The coordinate conversion $dv=\mathrm dt+e^{-\delta}\mathrm dr/N$ and $d\varrho=e^\delta \mathrm dr$ relates the background normalization of $v$ to the asymptotic time. A dot denotes $\partial_v$ on the horizon.

Introducing an expansion parameter $\lambda$ for the perturbation amplitude, we denote $\delta X = \partial_\lambda X|_0$ and let $X^{(n)}$ represent the coefficient of $\lambda^n$ (without factorials). The near-horizon fields are expanded as
\begin{equation}
 \begin{aligned}
 q_{AB}&=q^{(0)}_{AB}+\lambda\chi_{AB}+O(\lambda^2),\\
 q^{(0)}_{AB}&=\rh^2\Omega_{AB},\\
 \phi&=\phi_H+\lambda\psi+O(\lambda^2).
 \end{aligned}
 \label{eq:h-fields}
\end{equation}

Because these adapted coordinates dynamically track the perturbed event horizon, the geometric displacement of the horizon is intrinsically absorbed into the scalar perturbation $\psi$. Defining $F \equiv f_{,\phi,H}\psi = \delta f_H$, the first-order geometric deformation of the horizon is characterized by the expansion $\theta$, shear $\sigma_{AB}$, and perturbed normal connection $m_A$:
\begin{align}
	\chi &= q_{(0)}^{AB}\chi_{AB}, & B_{AB} &= \frac{1}{2}\dot\chi_{AB}, \nonumber\\
	\theta &= q_{(0)}^{AB}B_{AB}, & \sigma_{AB} &= B_{AB} - \frac{1}{2}\theta q^{(0)}_{AB}, \nonumber\\
	V_A &= \mathcal D^BB_{AB} - \mathcal D_A\theta, & m_A &= \frac{1}{2}\delta\beta_A.
	\label{eq:h-geometry}
\end{align}
Here, $\mathcal D_A$ is the covariant derivative associated with the background sphere $q^{(0)}_{AB}$, and background sphere indices are raised using $q_{(0)}^{AB}$. The intrinsic curvature of the background sphere and its first variation are
\begin{align}
	\mathcal R_0&=\frac{2}{\rh^2},\qquad
	\mathcal D^2=\mathcal D^A\mathcal D_A,\nonumber\\
	\delta\mathcal R&=\mathcal D_A\mathcal D_B\chi^{AB}
	-\mathcal D^2\chi-\tfrac12\mathcal R_0\chi.
	\label{eq:h-curvature}
\end{align}
The static transverse gradients required for the boundary response evaluate to
\begin{equation}
	c_H = \frac{e^{-\delta_H}}{\rh},\qquad
	d_H = \partial_\varrho f\big|_H = e^{-\delta_H}f_{,\phi,H}\phi'_H.
	\label{eq:h-transverse}
\end{equation}
The coefficient $c_H$ measures the transverse deformation of the background cut metric through $\partial_\varrho q_{AB}|_{\lambda=0,\varrho=0}=2c_Hq^{(0)}_{AB}$, while $d_H$ is the transverse gradient of the scalar coupling. These background quantities also satisfy $\mathcal A=1-4\alpha\kappa d_H$.

We place the stretched surface at a small positive coordinate value $\varrho$, where the lapse is $\varepsilon=\sqrt{2\kappa\varrho}$ to leading order. As $\varepsilon \to 0$, its observers become infinitely redshifted ($\varepsilon u\to\ell$ and $\varepsilon n\to\ell$). Regularity further requires $\varepsilon n^\mu\nabla_\mu\phi\to\dot\phi$. The finite horizon limits are extracted with the corresponding redshift weights:
\begin{align}
	\rho_H &= \lim_{\varepsilon\to0}\varepsilon\rho, 
	&s_{HAB} &= \lim_{\varepsilon\to0}\varepsilon s_{AB},\nonumber\\
	\mathcal P_A &= \lim_{\varepsilon\to0}(-t_{uA}), 
	&j_H &= \lim_{\varepsilon\to0}\varepsilon j_\phi,
	\label{eq:h-limits}
\end{align}
where $s_{AB}=t_{AB}$ and $t_{uA}=t_{ab}u^ae_A^b$ for tangential cut vectors $e_A^b$. The two-dimensional pressure is defined as $p_H=q^{AB}s_{HAB}/2$, leaving the traceless stress as $\tau_{AB}=s_{HAB}-p_Hq_{AB}$. The momentum projection $\mathcal P_A$ requires no additional $\varepsilon$ weighting to remain finite.

The amplitude variation must be taken before the horizon limit. For example, the normal coupling derivative contains both a vanishing background term and a divergent first variation:
\begin{equation}
	n^\mu\nabla_\mu f = \varepsilon d_H + \lambda\frac{\dot F}{\varepsilon} + O(\varepsilon^3) + \lambda O(\varepsilon) + O(\lambda^2).
	\label{eq:h-normal-orders}
\end{equation}
This order retains contributions from products of vanishing background factors and divergent first variations, following the treatment of the singular limit in Ref.~\cite{JMS2017}. All generator derivatives present at linear amplitude (including $\dot\sigma_{AB}$ and $\ddot\psi$) are retained, together with the full $\alpha$ dependence; a separate constitutive derivative expansion is used only to extract transport coefficients.

\subsection{Linear membrane response and its interpretation}
\label{sec:response}
\label{sec:linear-response}

On the static background, the regulated horizon responses evaluated in equilibrium are
\begin{equation}
 \begin{aligned}
 \rho_H^{(0)}&=0,\quad p_H^{(0)}=\frac\kappa{8\pi G},\\
 \tau_{AB}^{(0)}&=0,\quad\mathcal P_A^{(0)}=0,\\
 j_H^{(0)}&=\frac{\alpha\kappa f_{,\phi,H}\mathcal R_0}{4\pi G}.
 \end{aligned}
 \label{eq:h-equilibrium}
\end{equation}

Their first-order variations are obtained by inserting the near-horizon fields into Eqs.~\eqref{eq:m-stress} and \eqref{eq:m-scalar}, using the order of limits discussed above:
\begin{equation}
 \begin{aligned}
 8\pi G\,\delta\rho_H={}&-\theta-2\alpha\mathcal R_0\dot F,\\
 8\pi G\,\delta p_H={}&\frac{\mathcal A}{2}\theta+2\alpha d_H\dot\theta
                         -2\alpha\kappa\mathcal D^2F\\
 &+4\alpha c_H(\ddot F-\kappa\dot F),\\
 8\pi G\,\delta\tau_{AB}={}&-\mathcal A\sigma_{AB}-4\alpha d_H\dot\sigma_{AB}\\
 &+4\alpha\kappa(\mathcal D_A\mathcal D_BF)_{\rm TF},\\
 8\pi G\,\delta\mathcal P_A={}&m_A-4\alpha c_H\mathcal D_A\dot F
                             +4\alpha d_HV_A,\\
 4\pi G\,\delta j_H={}&-\dot\psi+\alpha\kappa\bigl(f_{,\phi,H}\delta\mathcal R
                         +f_{,\phi\phi,H}\mathcal R_0\psi\bigr)\\
 &+2\alpha c_Hf_{,\phi,H}(\dot\theta-\kappa\theta).
 \end{aligned}
 \label{eq:h-response}
\end{equation}
Here, the trace-free part of a symmetric cut tensor is defined as $X_{AB}^{\rm TF} \equiv X_{AB} - \frac{1}{2}q^{(0)}_{AB}q_{(0)}^{CD}X_{CD}$. Detailed derivations of these regulated contractions and the observer projections are given in Appendix~\ref{app:boundary}, especially Sec.~\ref{app:linear-details}. The pressure variation includes the variation of the inverse metric used in its trace, while the observer frame must also be varied when extracting the momentum response.

Equation~\eqref{eq:h-response} shows the coupled geometric and scalar response of the membrane. The surface energy density responds to both geometric expansion and scalar evolution. The angular Hessian of the scalar coupling contributes to the traceless spatial stress, while the intrinsic horizon-cut curvature and expansion enter the scalar response. The equilibrium scalar response~\eqref{eq:h-equilibrium} arises from the static curvature coupling even though the background scalar has no generator-time derivative.

To establish a fluid correspondence, we identify the shear ($\eta$) and bulk ($\zeta$) viscosities by extracting the first-derivative terms proportional to $\sigma_{AB}$ and $\theta$ from the spatial stress response ($\delta\tau_{AB} = -2\eta\sigma_{AB} + \dots$ and $\delta p_H = -\zeta\theta + \dots$):
\begin{equation}
	\eta = \frac{\mathcal A}{16\pi G},\qquad \zeta = -\frac{\mathcal A}{16\pi G}.
	\label{eq:e-viscosities}
\end{equation}
The shear viscosity is modulated by the geometric weight $\mathcal A$, consistent with the relative tensor kinetic coefficient found in spherical odd-parity perturbations~\cite{KMS2012}. The negative bulk viscosity is familiar for event-horizon membranes~\cite{PriceThorne1986,JMS2017}. The full response in Eq.~\eqref{eq:h-response} also includes metric-scalar mixing and higher generator derivatives. Section~\ref{sec:entropy} evaluates the work of these responses subject to the horizon constraints.

\subsection{Surface balances and horizon constraints}
\label{sec:finite-balances}

The metric and scalar responses obey a common boundary identity. Because both $h_{ab}$ and $\phi$ vary under a tangential diffeomorphism, the invariance of the on-shell boundary functional implies the momentum constraint~\cite{JulieBerti2020}
\begin{equation}
	D_at^a{}_b = j_\phi D_b\phi.
	\label{eq:m-ward}
\end{equation}
The right-hand side represents the force exerted by the scalar field on the surface fluid. Projecting this identity along the observer velocity $u^a$ yields the surface energy balance
\begin{equation}
 \begin{aligned}
 &u(\rho)+(\rho+p)\Theta+\mathcal D_a\mathcal Q^a
       +2a_a\mathcal Q^a\\
 &\quad+\Pi^{ab}\varsigma_{ab}=-j_\phi u(\phi).
 \end{aligned}
 \label{eq:m-finite-energy}
\end{equation}

Here, the directional derivative is $u(X) \equiv u^aD_aX$, the observer acceleration is $a_b \equiv u^aD_au_b$, the fluid expansion is $\Theta \equiv D_au^a$, and $\varsigma_{ab}$ denotes the observer shear. The spatial derivative $\mathcal D_a$ is projected onto the orthogonal spatial cuts. The terms $p\Theta$, $\Pi^{ab}\varsigma_{ab}$, and $j_\phi u(\phi)$ give the work of the isotropic pressure, traceless spatial stress, and scalar response, respectively. The derivation of this energy equation and its spatial momentum counterpart is detailed in Appendix~\ref{app:finite-balances}.

To extract the linear horizon limit, we expand the kinematic variables as $\Theta = \lambda\theta/\varepsilon + O(\lambda\varepsilon)$ and $u(\phi) = \lambda\dot\psi/\varepsilon + O(\lambda\varepsilon)$. Multiplying the energy balance by $\varepsilon^2$ and the spatial momentum balance by $\varepsilon$ isolates the leading-order horizon dynamics:
\begin{align}
	\partial_v\delta\rho_H + p_H^{(0)}\theta &= -j_H^{(0)}\dot\psi, \label{eq:h-energy} \\
	\partial_v\delta\mathcal P_A + \mathcal D_A\delta p_H + \mathcal D^B\delta\tau_{AB} &= j_H^{(0)}\mathcal D_A\psi. \label{eq:h-momentum}
\end{align}
The fixed-nonaffinity gauge removes the leading angular acceleration. Substituting the linear responses from Eq.~\eqref{eq:h-response} into these balances yields the horizon evolution equations:
\begin{align}
	(\partial_v-\kappa)(\theta+2\alpha\mathcal R_0\dot F) &= 0, \label{eq:h-null} \\
	\dot m_A - \mathcal A V_A - 4\alpha\kappa c_H\mathcal D_A\dot F &= 0. \label{eq:h-codazzi}
\end{align}
The cancellations of the spatial divergence terms are demonstrated in Appendix~\ref{app:linear-details}.

These balances can also be obtained from the purely null and null-spatial projections of the linearized metric field equations on the horizon. From the Raychaudhuri identity, we have $\delta G_{\ell\ell} = (\kappa-\partial_v)\theta$. Incorporating the scalar-Gauss-Bonnet contributions gives
\begin{equation}
	\begin{aligned}
		\delta E_{\ell\ell} &= (\kappa-\partial_v)(\theta+2\alpha\mathcal R_0\dot F), \\
		\delta E_{\ell A} &= -\dot m_A + \mathcal A V_A + 4\alpha\kappa c_H\mathcal D_A\dot F,
	\end{aligned}
	\label{eq:h-bulk}
\end{equation}
where $E_{\ell\ell} \equiv E_{\mu\nu}\ell^\mu\ell^\nu$ and $E_{\ell A} \equiv E_{\mu\nu}\ell^\mu e_A^\nu$ for a vector $e_A^\nu$ tangent to the spatial horizon cut. The membrane energy and momentum balances therefore reproduce $\delta E_{\ell\ell}=0$ and $\delta E_{\ell A}=0$ at linear order.

Finally, we consider an event horizon that asymptotically settles to a stationary state. Equation~\eqref{eq:h-null} admits a homogeneous solution $C(x)e^{\kappa v}$ for the combined rate $\theta+2\alpha\mathcal R_0\dot F$. Because the surface gravity is positive ($\kappa>0$), future stationarity excludes this exponentially growing mode and gives
\begin{equation}
 \begin{aligned}
 \theta+2\alpha\mathcal R_0\dot F&\longrightarrow0
                          \quad(v\longrightarrow+\infty),\\
 \theta&=-2\alpha\mathcal R_0\dot F.
 \end{aligned}
 \label{eq:h-future}
\end{equation}

This teleological condition relates the geometric expansion to the scalar rate on the event horizon. Section~\ref{sec:entropy} uses this relation to evaluate the combined work, determine the horizon absorption, and derive the entropy balance.

For the slow growing and decaying modes studied in
Sec.~\ref{sec:dynamical-input}, the same relation follows directly from
Eq.~\eqref{eq:h-null}, since their growth or decay rate differs from
$\kappa$.

\section{Horizon absorption and entropy balance}
\label{sec:entropy}

The combined work of the membrane stress and scalar response contains
equilibrium, storage, and absorption contributions. In EsGB gravity,
scalar evolution also induces horizon expansion at first order through
the curvature coupling (Eq.~\eqref{eq:h-future}), so the absorbed power
includes both scalar work and expansion work. We separate these
contributions and relate the absorption to horizon entropy evolution.
Between stationary cuts, the integrated absorption equals the horizon
temperature times the quadratic entropy change; for arbitrary cuts,
the balance also contains entropy-rate and storage endpoint terms.

\subsection{Membrane transport and equilibrium entropy}

The membrane viscosities in Eq.~\eqref{eq:e-viscosities} identify
the first-derivative shear and expansion response. To relate the full
coupled work to entropy, we first specify the entropy of stationary
horizon cuts and its extension away from stationarity.

For a stationary cut, the Wald entropy is~\cite{Wald1993,JulieBerti2019}
\begin{equation}
 S_W=\frac{1}{4G}\int_H \mathrm d^2x\sqrt q
       \left(1+2\alpha f_H\mathcal R[q]\right).
 \label{eq:m-wald-integral}
\end{equation}
For a round cut, $\mathcal R=2/\rh^2$ and
$\int_H\sqrt q\,\mathcal R=8\pi$, giving
\begin{equation}
 \begin{aligned}
 A_H&=4\pi\rh^2,\qquad T_0=\frac\kappa{2\pi},\\
 S_W&=\frac{\pi\rh^2}{G}+\frac{4\pi\alpha f_H}{G}.
 \end{aligned}
 \label{eq:m-entropy}
\end{equation}

To distinguish changes of equilibrium from irreversible work, consider
the entropy associated with the intrinsic geometry and scalar on a
closed horizon cut. We use
\begin{equation}
 \begin{aligned}
 S_{\rm intr}[q,\phi]&=\frac1{4G}\int\mathrm d^2x\sqrt q
                      [1+2\alpha f(\phi)\mathcal R],\\
 I_{\rm eq}&=T_0S_{\rm intr},\qquad T_0=\frac\kappa{2\pi}.
 \end{aligned}
 \label{eq:e-intrinsic}
\end{equation}

On a stationary cut $S_{\rm intr}$ agrees with the Wald entropy.
For evolving cuts this is an explicit choice of extension, relevant
to the endpoint terms in a finite-cut balance \cite{JKM1994}.
We hold the background temperature $T_0$ fixed when varying $I_{\rm eq}$.
This functional identifies the work associated with the instantaneous
intrinsic state; its use does not assume that an evolving cut is
itself in equilibrium.
The future-horizon condition also fixes the first-order entropy
rate on a closed cut:
\begin{equation}
 \partial_v S_{\rm intr}^{(1)}
 =\frac1{4G}\int\mathrm dA_0\,
           (\theta+2\alpha\mathcal R_0\dot F)=0 .
 \label{eq:h-linear-entropy}
\end{equation}
Here $\mathrm dA_0=\sqrt{q^{(0)}}\mathrm d^2x$ is the background
area element, and the angular divergence in the curvature variation
integrates to zero. The nonzero entropy production considered below begins at quadratic order.

To compute the intrinsic functional, we need the variation of the two-dimensional curvature. The identity $\mathcal R_{AB}=\tfrac12\mathcal R q_{AB}$ in two dimensions makes this variation particularly simple. Let $\mathcal D_A$ denote the connection of the full nonlinear cut metric (which reduces to the background connection when evaluated on the background). On a closed cut, we have
\begin{equation}
 \begin{aligned}
 &\delta\int\mathrm d^2x\sqrt q\,f\mathcal R\\
 &\quad=\int\mathrm d^2x\sqrt q\,
       \Bigl[(\mathcal D^A\mathcal D^Bf-q^{AB}\mathcal D^2f)
                    \delta q_{AB}\\
 &\hspace{43mm}+f_{,\phi}\mathcal R\,\delta\phi\Bigr].
 \end{aligned}
 \label{eq:e-curvature-variation}
\end{equation}

There is no term proportional to
$f(\mathcal R^{AB}-\tfrac12q^{AB}\mathcal R)$ in two dimensions.
Adding the area variation gives the two equilibrium responses
conjugate to $q_{AB}$ and $\phi$:
\begin{align}
 \delta I_{\rm eq}
 &=\int \mathrm d^2x\sqrt q\,
   \left(\frac12s_{\rm eq}^{AB}\delta q_{AB}
                          +j_{\rm eq}\delta\phi\right),\nonumber\\
 s_{{\rm eq}\,AB}
 &=\frac\kappa{8\pi G}\left[
 q_{AB}+4\alpha(\mathcal D_A\mathcal D_Bf
                          -q_{AB}\mathcal D^2f)\right],\nonumber\\
 j_{\rm eq}&=\frac{\alpha\kappa}{4\pi G}
                         f_{,\phi}\mathcal R.
 \label{eq:e-equilibrium}
\end{align}
These expressions reproduce the spatial Hessian and intrinsic-curvature
terms already present in the membrane response. The comparison must be
made at the level of the canonical densities
$\sqrt q\,s^{AB}/2$ and $\sqrt q\,j$, including the change in
the area element. In particular, the trace contribution from varying
the cut curvature cancels the corresponding area term; remaining
angular divergences combine with the Hessian stress after integration
by parts. The functional $I_{\rm eq}=T_0S_{\rm intr}$ therefore
accounts for the work associated with an instantaneous change of the
intrinsic horizon state.

One sometimes characterizes horizon transport by the ratio of shear
viscosity to entropy density. For the spherical background, the
stationary Wald entropy in Eq.~\eqref{eq:m-entropy} gives
$s_W=S_W/A_H=(1+4\alpha f_H/\rh^2)/(4G)$, hence, when $s_W\ne0$,
\begin{equation}
 \frac{\eta}{s_W}
 =\frac{\mathcal A}{4\pi(1+4\alpha f_H/\rh^2)}.
 \label{eq:e-eta-s}
\end{equation}
This uses the Wald entropy of the specified action
\cite{Wald1993,JulieBerti2019}. In four dimensions the replacement
$f\mapsto f+C$ adds a topological term to the action. It shifts the
entropy of a spherical horizon by $4\pi\alpha C/G$ without changing
the field equations or the membrane transport \cite{Liko2008}.
Thus the absolute entropy normalization matters for the ratio, although
not for the entropy difference between cuts of fixed topology.

\subsection{Coupled work and horizon absorption}

Subtracting this equilibrium response from the linear membrane
response leaves terms containing the rates of change of the horizon.
Denote the remainder of the pressure, traceless stress, and scalar
response by $\delta p_{\rm neq}$,
$\delta\tau_{{\rm neq}\,AB}$, and $\delta j_{\rm neq}$.
The parts that contribute to quadratic work are
\begin{align}
 \delta p_{\rm neq}={}&\frac1{8\pi G}\Bigl[
        \frac{\mathcal A}{2}\theta+2\alpha d_H\dot\theta\nonumber\\
 &+4\alpha c_Hf_{,\phi,H}(\ddot\psi-\kappa\dot\psi)\Bigr],\nonumber\\
 \delta\tau_{{\rm neq}\,AB}={}&\frac1{8\pi G}
        [-\mathcal A\sigma_{AB}-4\alpha d_H\dot\sigma_{AB}],\nonumber\\
 \delta j_{\rm neq}={}&\frac1{4\pi G}\Bigl[-\dot\psi\nonumber\\
 &+2\alpha c_Hf_{,\phi,H}(\dot\theta-\kappa\theta)\Bigr].
 \label{eq:e-nonequilibrium}
\end{align}

The curvature coupling introduces scalar time derivatives into the
pressure and expansion derivatives into the scalar response.
The second generator derivatives contribute to total time derivatives
in the work and hence to the balance between finite horizon cuts.

A first-order stress or scalar response does work against a first-order
rate at second order in perturbation amplitude. These work contractions
enter with a minus sign in the membrane energy equation,
Eq.~\eqref{eq:m-finite-energy}; absorption contributes negatively
to the work contraction.
Contracting the traceless stress with the shear, multiplying the pressure by the
expansion, and multiplying the scalar response by $\dot\psi$,
we obtain, with $\sigma^2=\sigma_{AB}\sigma^{AB}$,
\begin{align}
 &\delta p_{\rm neq}\theta+\delta\tau_{\rm neq}^{AB}\sigma_{AB}
                    +\delta j_{\rm neq}\dot\psi\nonumber\\
 &\quad=\frac1{8\pi G}\Bigl[\frac{\mathcal A}{2}\theta^2
                    -\mathcal A\sigma^2-2\dot\psi^2\nonumber\\
 &\hspace{20mm}-8\alpha\kappa c_Hf_{,\phi,H}\theta\dot\psi\Bigr]
                    +\partial_v\mathcal K_2,
 \label{eq:e-work}
\end{align}

where the quantity whose time derivative appears is
\begin{equation}
 \mathcal K_2
 =\frac{\alpha d_H\theta^2-2\alpha d_H\sigma^2
           +4\alpha c_Hf_{,\phi,H}\theta\dot\psi}{8\pi G}.
 \label{eq:e-storage}
\end{equation}
The time integral of $\partial_v\mathcal K_2$ depends only on the
values of $\mathcal K_2$ at the two cuts. We refer to this
total-time-derivative contribution to the work balance as temporal
storage. It need not have a definite sign and is distinct from the
absorbed power. The relative signs are fixed by the complete
work contraction; notably, the shear term enters $\mathcal K_2$
with a minus sign. The derivative terms combine by the identities
\begin{align}
 2\alpha d_H\theta\dot\theta
       &=\partial_v(\alpha d_H\theta^2),\nonumber\\
 -4\alpha d_H\sigma^{AB}\dot\sigma_{AB}
       &=\partial_v(-2\alpha d_H\sigma^2),\nonumber\\
 4\alpha c_Hf_{,\phi,H}
       (\theta\ddot\psi+\dot\theta\dot\psi)
       &=\partial_v(4\alpha c_Hf_{,\phi,H}\theta\dot\psi).
 \label{eq:e-product-rules}
\end{align}
The coefficients $c_H$ and $d_H$ are constant along each generator
of the static background. After removing the total derivative, the
remaining work still contains
$-8\alpha\kappa c_Hf_{,\phi,H}\theta\dot\psi$.
This mixed term has the same quadratic order as the apparent shear,
expansion, and scalar losses. Its physical interpretation depends on
the allowed relation between $\theta$ and $\dot\psi$.

For the future event horizon that settles to a stationary state,
Eq.~\eqref{eq:h-future} provides precisely that relation:
\begin{equation}
 \theta=-2\alpha\mathcal R_0f_{,\phi,H}\dot\psi,
 \qquad \mathcal R_0=\frac2{\rh^2}.
 \label{eq:e-constraint}
\end{equation}
The relation holds pointwise on the perturbed sphere after the
future-growing homogeneous solution has been excluded.
Section~\ref{sec:dynamical-input} shows directly from the linear horizon
constraint that the slow growing and decaying modes considered there
obey the same relation. Substituting it into the complete work reduces
the coefficient of $\dot\psi^2$ according to
\begin{align}
 &1-\alpha^2 f_{,\phi,H}^{\,2}
       \Bigl(\mathcal A\mathcal R_0^2
                         +8\kappa c_H\mathcal R_0\Bigr)\nonumber\\
 &\quad=1-\frac{12\alpha^2 f_{,\phi,H}^{\,2}\mathcal A}{\rh^4}
                         =\Delta_H.
 \label{eq:e-discriminant}
\end{align}

We used $\kappa c_H=\mathcal A/(2\rh^2)$ together with the
regularity relation $\mathcal A=2/(1+\Delta_H)$. The work is thus
$-\mathcal P_2+\partial_v\mathcal K_2$, where
\begin{equation}
 \mathcal P_2
 =\frac{\mathcal A}{8\pi G}\sigma^2
            +\frac{\Delta_H}{4\pi G}\dot\psi^2.
 \label{eq:e-power}
\end{equation}
Here $\mathcal P_2$ is the quadratic coefficient of the absorbed power
per unit background area; the total absorbed power is
$\lambda^2\int\mathrm dA_0\,\mathcal P_2+O(\lambda^3)$.
It contains two nonnegative contributions on the regular
branch, since $\mathcal A>0$ and $\Delta_H>0$. The shear weight
$\mathcal A$ is visible in the independent constitutive response.
The scalar weight $\Delta_H$ is different: it emerges only after the
scalar work has been combined with the work of the induced expansion.

\subsection{Entropy balance between horizon cuts}

We now ask how the power in Eq.~\eqref{eq:e-power} is recorded in
horizon entropy. There is a subtlety in taking the null limit:
the density and spatial flux on a stretched membrane can contain
terms singular in its lapse, although their cut-integrated balance
has a finite perturbative limit. We therefore integrate the energy
equation over a closed cut while the surface is still timelike.
In the same Gaussian null coordinates used in Sec.~\ref{sec:model},
write the induced metric on the stretched surface as
$h_{ab}\mathrm dx^a\mathrm dx^b=-\mathcal N^2\mathrm dv^2+
q_{AB}(\mathrm dx^A+w^A\mathrm dv)(\mathrm dx^B+w^B\mathrm dv)$.
Here $\mathcal N$ and $w^A$ are its lapse and spatial shift.
Equation~\eqref{eq:h-coordinates} gives
$\mathcal N^2=2\kappa\varrho+O(\varrho^2)
=\varepsilon^2+O(\varrho^2)$ and
$w^A=\varrho q^{AB}\beta_B+O(\varrho^2)$.
Thus $\mathcal N$ reduces to the earlier redshift parameter at leading
order, while $w^A$ vanishes in the horizon limit.
Define $\mathcal E_{\rm mem}=\int\mathrm d^2x\sqrt q\,\mathcal N\rho$;
the observers normal to the spatial cuts have
$u=\mathcal N^{-1}(\partial_v-w^A\partial_A)$.
Multiplying Eq.~\eqref{eq:m-finite-energy} by
$\sqrt q\,\mathcal N^2$ turns the combined energy-flux and
acceleration terms into a spatial divergence,
$\mathcal D_A(\mathcal N^2\mathcal Q^A)$, whose integral vanishes.
The resulting cut balance is
\begin{align}
 \partial_v\mathcal E_{\rm mem}
 ={}&-\int\mathrm d^2x\sqrt q\,\mathcal N^2\nonumber\\
 &\quad\times(p\Theta+\Pi^{AB}\varsigma_{AB}+j_\phi u(\phi))\nonumber\\
 &+\int\mathrm d^2x\sqrt q\,\mathcal N\rho\,
            (\partial_v-w^A\mathcal D_A)\ln\mathcal N.
 \label{eq:e-regulated-balance}
\end{align}

This order of operations cancels possible singular spatial-flux
terms before the regulator is removed. In the fixed-nonaffinity
gauge, the lapse-gradient term in the second line is
$O(\mathcal N^2)$ times the density factor and does not survive
through quadratic perturbative order.

The remaining membrane energy can be related to the rate of the
intrinsic entropy functional. Let
$\mathcal B_{AB}=\dot q_{AB}/2$ and
$\vartheta=q^{AB}\mathcal B_{AB}$ for the full, not yet linearized,
horizon deformation. The tensor $\mathcal B_{AB}$ includes both
expansion and shear; its trace is $\vartheta$. Their relation to the
linear quantities of Sec.~\ref{sec:model} is
$\mathcal B_{AB}=\lambda B_{AB}+O(\lambda^2)$ and
$\vartheta=\lambda\theta+O(\lambda^2)$.
Differentiating the intrinsic curvature gives
\begin{equation}
 \dot{\mathcal R}=-\mathcal R\vartheta
       +2\mathcal D_A\mathcal D_B\mathcal B^{AB}
       -2\mathcal D^2\vartheta.
 \label{eq:e-curvature-rate}
\end{equation}
After integrating the angular derivatives by parts twice,
\begin{align}
 \dot S_{\rm intr}
 ={}&\frac1{4G}\int\mathrm d^2x\sqrt q\,
          \Bigl[\vartheta+2\alpha\mathcal R\dot f\nonumber\\
 &+4\alpha(\mathcal B^{AB}\mathcal D_A\mathcal D_Bf
                         -\vartheta\mathcal D^2f)\Bigr].
 \label{eq:e-full-entropy-rate}
\end{align}

The terms algebraic in the curvature cancel between the change of
the area element and the change of the two-dimensional Ricci scalar.
A direct expansion of the boundary energy density gives the same
bracket as Eq.~\eqref{eq:e-full-entropy-rate}, with coefficient
$-1/(8\pi G)$. Appendix~\ref{app:entropy} shows this cancellation,
including the transverse-gradient terms. The remaining pole contains
three first-order rates and begins at
$O(\lambda^3/\mathcal N^2)$; because the perturbation coefficients
are taken before the lapse limit, it does not enter at orders zero,
one, or two. Hence
\begin{equation}
 \lim_{\varepsilon\to0}\mathcal E_{\rm mem}^{(n)}
       =-\frac1{2\pi}\partial_vS_{\rm intr}^{(n)},
       \qquad n=0,1,2.
 \label{eq:e-density-storage}
\end{equation}

The superscript $(2)$ denotes the coefficient of
$\lambda^2$, so that
$X^{(2)}=\tfrac12\partial_\lambda^2X|_{\lambda=0}$.
With this convention, write
$q_{AB}=q^{(0)}_{AB}+\lambda\chi_{AB}
+\lambda^2q^{(2)}_{AB}+\cdots$ and
$\phi=\phi_H+\lambda\psi+\lambda^2\psi^{(2)}+\cdots$.
Their measure and coupling expansions include
\begin{align}
 \frac{\sqrt q}{\sqrt{q^{(0)}}}
 ={}&1+\frac\lambda2\chi\nonumber\\
 &+\lambda^2\Bigl(\tfrac12q_{(0)}^{AB}q^{(2)}_{AB}
                    +\tfrac18\chi^2\nonumber\\
 &\hspace{18mm}-\tfrac14\chi_{AB}\chi^{AB}\Bigr)
                    +O(\lambda^3),\nonumber\\
 f={}&f_H+\lambda f_{,\phi,H}\psi\nonumber\\
 &+\lambda^2\left(f_{,\phi,H}\psi^{(2)}
       +\tfrac12f_{,\phi\phi,H}\psi^2\right)+O(\lambda^3).
 \label{eq:e-second-order-measure}
\end{align}

The entropy $S_{\rm intr}^{(2)}$ includes the second-order metric and
scalar fields, their contribution to the intrinsic curvature, and
quadratic combinations of first-order fields. The quadratic work and
storage depend on the first-order responses and rates.

The equilibrium functional contributes
$T_0\dot S_{\rm intr}$ to the work. The remainder is
$-\mathcal P_2+\partial_v\mathcal K_2$.
Combining these facts with the integrated membrane energy equation
gives, with $K_2(v)=\int\mathrm dA_0\,\mathcal K_2$,
\begin{equation}
 (\partial_v-\kappa)\partial_vS_{\rm intr}^{(2)}
  =-2\pi\int\mathrm dA_0\,\mathcal P_2+2\pi\partial_vK_2.
 \label{eq:e-differential}
\end{equation}
The second-order fields are contained on the left through the
entropy rate. The power and storage on the right are quadratic
in first-order fields, so their integrals use the background
area measure.

Integrating between arbitrary cuts $v_i$ and $v_f$ yields
\begin{equation}
 \begin{aligned}
 T_0[S_{\rm intr}^{(2)}]_{v_i}^{v_f}
 ={}&\int_{v_i}^{v_f}\mathrm dv\int\mathrm dA_0\,\mathcal P_2\\
 &+\frac1{2\pi}[\dot S_{\rm intr}^{(2)}]_{v_i}^{v_f}
             -[K_2]_{v_i}^{v_f}.
 \end{aligned}
 \label{eq:e-finite-cuts}
\end{equation}

Here $[X]_{v_i}^{v_f}=X(v_f)-X(v_i)$. In addition to the integrated
absorption, Eq.~\eqref{eq:e-finite-cuts} contains endpoint differences
in the entropy rate and temporal storage. The entropy-rate term comes
from the second time derivative of the entropy in
Eq.~\eqref{eq:e-differential}, whereas the storage term comes from
the total time derivative in the coupled work. The future boundary
condition selecting the event horizon allows a nonzero entropy rate
on a cut before a later pulse reaches the horizon.
Similar endpoint terms appear in higher-curvature physical-process
analyses~\cite{Mishra2018}.

For a process that approaches stationary horizons in the distant
past and future, the rates of the first-order fields,
$\dot S_{\rm intr}^{(2)}$, and the temporal storage vanish at the
endpoints. The intrinsic functional there equals the stationary
Wald entropy. Equation~\eqref{eq:e-finite-cuts} then becomes
\begin{equation}
 \begin{aligned}
 T_0\Delta S_W^{(2)}
 ={}&\int_{-\infty}^{\infty}\!\mathrm dv\,\int\!\mathrm dA_0\,\\
 &\left[\frac{\mathcal A}{8\pi G}\sigma^2
             +\frac{\Delta_H}{4\pi G}\dot\psi^2\right]\geq0.
 \end{aligned}
 \label{eq:e-law}
\end{equation}

Equation~\eqref{eq:e-law} gives the entropy change between stationary
cuts as a positive quadratic integral. Appendix~\ref{app:entropy}
checks the scalar power and storage independently, using the bulk
null equation in spherical symmetry.

On the regular branch, the absorption weights satisfy
$0<\Delta_H\leq1$ and $1\leq\mathcal A<2$.
The negative bulk viscosity in Eq.~\eqref{eq:e-viscosities} is
therefore compatible with the positive power: expansion has already
been linked to the scalar rate in the physical event-horizon
perturbation.

As a normalization check, set $\alpha=0$ while retaining the
minimally coupled scalar. The linear responses become
\begin{align}
 \delta\rho_H&=-\frac{\theta}{8\pi G},&
 \delta p_H&=\frac{\theta}{16\pi G},\nonumber\\
 \delta\tau_{AB}&=-\frac{\sigma_{AB}}{8\pi G},&
 \delta j_H&=-\frac{\dot\psi}{4\pi G}.
 \label{eq:e-einstein-limit}
\end{align}
Then the temporal storage in Eq.~\eqref{eq:e-storage} vanishes and
the future-horizon constraint gives $\theta=0$ at first order.
The constrained power reduces to
$\sigma^2/(8\pi G)+\dot\psi^2/(4\pi G)$; the scalar term is absent
only when the scalar field itself is not perturbed.

Quadratic horizon entropy balances are also studied in constant-coupling
Lovelock gravity \cite{FGS2018} and in connection with canonical energy
\cite{HWZ2024}. For four-dimensional scalar--Gauss--Bonnet gravity,
Hollands, Kov\'acs and Reall obtained an entropy construction in an
effective-field-theory expansion \cite{HKR2022}.
Writing their EFT scale as $\ell_{\rm EFT}$, scalar as
$\Phi_{\rm EFT}$, and coupling as $\beta_{\rm EFT}$, their
conventions map to ours through $\Phi_{\rm EFT}=2\phi$ and
$\ell_{\rm EFT}^2\beta_{\rm EFT}(\Phi_{\rm EFT})
=4\alpha f(\phi)$. The resulting intrinsic entropy density agrees
with Eq.~\eqref{eq:e-intrinsic}. Their explicit
scalar--Gauss--Bonnet result retains an $O(\ell_{\rm EFT}^4)$
remainder, whereas $\mathcal A-1$ and $\Delta_H-1$ first occur at
$O(\alpha^2)$ under this identification. Our calculation gives the
corresponding exact-coupling weights within the specified EsGB
action, the regular spherical branch, and the stated perturbative
orders; it does not remove the regime restrictions of an EFT
interpretation.

In Sec.~\ref{sec:slow}, the scalar absorption weight determines the
horizon energy loss of a slow spherical mode near a simple turning
point. For these modes, the additional time derivatives in
Eq.~\eqref{eq:e-differential} are subleading to the absorption term.
Combining this horizon loss with outgoing scalar radiation and the
effective force obtained from a static family at fixed Wald entropy
gives the leading growth or decay rate.

\section{Slow spherical dynamics near a simple turning point}
\label{sec:slow}

Black holes in EsGB gravity can exhibit critical phenomena associated
with the stability of scalar perturbations, such as spontaneous
scalarization. This section studies slow dynamics near a simple turning
point of a family of static, spherically symmetric black holes.
Specifically, we select a static black hole near the turning point,
hold the scalar field at infinity fixed, and study the growth or decay
of small spherical perturbations. To calculate the evolution rate,
we use a static family at fixed Wald entropy $S_W$ and coupling constant
$\alpha$, parametrized by the horizon scalar value $\Phi$.
Along this family, the black-hole mass $M$, scalar charge $Q$, and
asymptotic scalar value $\phi_\infty$ depend on $\Phi$:
\[
 M=M(\Phi),\qquad Q=Q(\Phi),\qquad
 \phi_\infty=\phi_\infty(\Phi).
\]
We consider a simple turning point of the curve $\phi_\infty(\Phi)$
at $\Phi_c$, where
\[
 \phi_{\infty,\Phi}(\Phi_c)=0,\qquad
 \phi_{\infty,\Phi\Phi}(\Phi_c)\neq0.
\]
A subscript ${,\Phi}$ denotes differentiation with respect to $\Phi$
along the static family at fixed $S_W$ and $\alpha$.

At the turning point, a small change in the horizon scalar value
leaves the asymptotic scalar value unchanged at first order.
The infinitesimal difference between neighboring static solutions
therefore gives a spherical static perturbation satisfying the boundary
condition with a fixed scalar value at infinity. On nearby black-hole
backgrounds, the corresponding perturbation can grow or decay slowly.
Our aim is to determine its leading exponential growth or decay rate
near the turning point.

We assume that a spherical mode near the turning point evolves on a
time scale much longer than the characteristic black-hole time scale,
and that the other, faster modes have relaxed. In the slow-dynamics
approximation, we use infinitesimal differences between neighboring
static black holes to describe the leading perturbation profile.
We also construct an effective potential describing the potential-energy
change associated with departures from equilibrium along this mode
at fixed external scalar source. Its negative derivative with respect
to the amplitude gives the effective force, which either restores
equilibrium or drives the perturbation further away from it.
Horizon absorption and outgoing scalar radiation provide two channels
for energy loss. Combining this force with the membrane absorption
power in Eq.~\eqref{eq:e-power} and the scalar radiation loss gives
the leading exponential growth or decay rate.

\subsection{Static family and effective potential}
\label{sec:amplitude}
\label{sec:static-force}

To describe the time evolution of the perturbation, we choose the scalar
field evaluated on the evolving event horizon as the amplitude variable:
\begin{equation}
 \Phi(t)=\phi\big|_{H(t)}
 =\phi_H+\lambda\psi(t)+O(\lambda^2).
 \label{eq:d-amplitude}
\end{equation}
Here $\lambda$ is the perturbation parameter in Eq.~\eqref{eq:h-fields},
and $\psi$ is the first-order scalar perturbation on the event horizon.
Since the field is evaluated on the evolving horizon, $\psi$ includes
the effect of the horizon displacement on the scalar value.
With our variation convention, $\delta\Phi=\psi$.

We describe the slow amplitude evolution using $t$, with the same
normalization as the Killing time at infinity. On the horizon, we
identify this time label with $v$, so that
\begin{equation}
 \frac{\mathrm d\Phi}{\mathrm dt}
 =\lambda\dot\psi+O(\lambda^2),
 \label{eq:d-amplitude-rate}
\end{equation}
where a dot denotes differentiation with respect to the horizon time
$v$. In the discussion of outgoing scalar radiation below, we denote
the exterior static time coordinate by $t_{\rm ext}$ and use the same
time normalization.

We determine the effective force from the static family at fixed Wald
entropy $S_W$ and coupling constant $\alpha$ introduced above.
The first-order entropy variation of the mode vanishes, and the entropy
change begins at second order. The leading effective force can therefore
be calculated along the static family at fixed Wald entropy.

The black-hole mass $M$, asymptotic scalar value $\phi_\infty$, and
scalar charge $Q$ are defined by the large-radius behavior
\begin{equation}
 \begin{aligned}
 N&=1-\frac{2GM}{r}+O(r^{-2}),\\
 \phi&=\phi_\infty+\frac{Q}{r}+O(r^{-2}).
 \end{aligned}
 \label{eq:d-asymptotics}
\end{equation}
The charge $Q$ has dimensions of length. In the dynamical problem,
the external scalar source $\phi_{\rm ext}$ at infinity is held fixed.
A static black hole is an equilibrium under this boundary condition
when
\[
 \phi_\infty(\Phi)=\phi_{\rm ext}.
\]

At fixed external source $\phi_{\rm ext}$, the effective potential
$E_{\rm eff}(\Phi;\phi_{\rm ext})$ describes the potential-energy
change as the amplitude $\Phi$ departs from equilibrium. Its negative
derivative with respect to $\Phi$ gives the effective force driving
the amplitude evolution:
\[
 F_{\rm eff}
 =-\left.\frac{\partial E_{\rm eff}}{\partial\Phi}
   \right|_{\phi_{\rm ext}}.
\]
Constructing this potential requires including the coupling of the
scalar field to the external source. Along the static family,
$\phi_\infty$ changes with $\Phi$, so the mass function $M(\Phi)$
describes static solutions with different source values. We use the
black-hole first law to express this energy relation in a form suited
to a fixed external source.

The required thermodynamic relation is the black-hole first law at
fixed coupling~\cite{JulieBerti2019}:
\begin{equation}
 \begin{aligned}
 \mathrm dM
 &=T\,\mathrm dS_W-\frac{Q}{G}\,\mathrm d\phi_\infty,\\
 M_{,\Phi}
 &=-\frac{Q}{G}\phi_{\infty,\Phi}
 \qquad(S_W,\alpha\ \text{fixed}).
 \end{aligned}
 \label{eq:d-first-law}
\end{equation}
Here $T$ is the Hawking temperature of each static member of the
family, and $T_0$ is the temperature of the chosen background.

We assume $Q_{,\Phi}\ne0$ near the turning point, so that the scalar
charge $Q$ also serves as a local parameter along the static family.
The first law gives
\[
 \mathrm d\left(M+\frac{\phi_\infty Q}{G}\right)
 =\frac{\phi_\infty}{G}\,\mathrm dQ
 \qquad(S_W,\alpha\ \text{fixed}).
\]
The combination $M+\phi_\infty Q/G$ thus expresses the static energy
as a function of scalar charge, corresponding to the Legendre transform
from $\phi_\infty$ to $Q$. Adding the interaction energy
$-\phi_{\rm ext}Q/G$ with the prescribed external source gives,
following Ref.~\cite{Khalil2019}, the effective potential for the slow
mode:
\begin{equation}
 E_{\rm eff}(\Phi;\phi_{\rm ext})
 =M(\Phi)
 +\frac{[\phi_\infty(\Phi)-\phi_{\rm ext}]Q(\Phi)}{G}.
 \label{eq:d-energy}
\end{equation}

Differentiating with respect to $\Phi$ at fixed $\phi_{\rm ext}$ gives
\begin{equation}
 \begin{aligned}
 E_{{\rm eff},\Phi}
 &=M_{,\Phi}
 +\frac{Q}{G}\phi_{\infty,\Phi}
 +\frac{Q_{,\Phi}}{G}
 [\phi_\infty-\phi_{\rm ext}]\\
 &=\frac{Q_{,\Phi}}{G}
 [\phi_\infty(\Phi)-\phi_{\rm ext}],
 \end{aligned}
 \label{eq:d-force}
\end{equation}
where Eq.~\eqref{eq:d-first-law} cancels the first two terms.
The generalized force is therefore
$F_{\rm eff}=-Q_{,\Phi}[\phi_\infty(\Phi)-\phi_{\rm ext}]/G$,
where $Q_{,\Phi}$ is the derivative of the scalar charge with respect
to $\Phi$ along the fixed-entropy family.

At an equilibrium $\Phi_0$ satisfying
$\phi_\infty(\Phi_0)=\phi_{\rm ext}$, the effective force vanishes.
For small departures from equilibrium, the potential curvature is
\begin{equation}
 E_{{\rm eff},\Phi\Phi}(\Phi_0)
 =\left.
 \frac{Q_{,\Phi}\phi_{\infty,\Phi}}{G}
 \right|_{\Phi_0},
 \label{eq:d-potential-curvature}
\end{equation}
and the corresponding linear effective force is
\[
 F_{\rm eff}
 =-E_{{\rm eff},\Phi\Phi}(\Phi_0)(\Phi-\Phi_0)
 +O\!\left((\Phi-\Phi_0)^2\right).
\]
Positive curvature gives a restoring force, whereas negative curvature
gives a destabilizing force. When the equilibrium lies at the turning
point, $\phi_{\infty,\Phi}=0$ and the potential curvature vanishes.
The linear restoring or destabilizing force is then zero, consistent
with the time-independent spherical perturbation discussed above.

The static family therefore determines the effective force on the slow
mode and the potential curvature near equilibrium.

\subsection{Horizon absorption and outgoing scalar radiation}
\label{sec:dynamical-input}

Having determined the effective force, we calculate the two energy-loss
channels associated with a changing amplitude $\Phi$: absorption by
the black hole and scalar radiation carrying energy to infinity.

We first calculate horizon absorption. Consider a perturbation with
generator-time dependence $e^{sv}$, where $s$ is its growth or decay
rate. Substitution into the linear horizon constraint,
Eq.~\eqref{eq:h-null}, gives
\[
 (s-\kappa)
 \left(\theta+2\alpha\mathcal R_0\dot F\right)=0.
\]
Here $F=f_{,\phi,H}\psi$ is the first-order perturbation of the
coupling function on the horizon. For a slow mode satisfying
$|s|\ll\kappa$,
\[
 \theta=-2\alpha\mathcal R_0\dot F.
\]
Applying this constraint to a spherical slow mode, whose horizon shear
vanishes, reduces the quadratic absorption power density in
Eq.~\eqref{eq:e-power} to
\[
 \left.\mathcal P_2\right|_{\sigma_{AB}=0}
 =\frac{\Delta_H}{4\pi G}\dot\psi^2.
\]
Using Eq.~\eqref{eq:d-amplitude-rate} and integrating over the background
horizon area gives
\begin{equation}
 \begin{aligned}
 P_H
 &=\lambda^2\int\mathrm dA_0\,
 \left.\mathcal P_2\right|_{\sigma_{AB}=0}
 +O(\lambda^3)\\
 &=\lambda^2\frac{r_H^2\Delta_H}{G}\dot\psi^2
 +O(\lambda^3)\\
 &=\frac{r_H^2\Delta_H}{G}
 \left(\frac{\mathrm d\Phi}{\mathrm dt}\right)^2
 +O(\lambda^3).
 \end{aligned}
 \label{eq:d-horizon-power}
\end{equation}
This is the power absorbed by the black hole as the amplitude $\Phi$
changes. The coefficient $r_H^2\Delta_H/G$ includes both the direct
scalar work and the work of the expansion induced by the scalar field.

The energy balance in Eq.~\eqref{eq:e-differential} also contains
$\partial_vK_2$ and $\partial_v^2S_{\rm intr}^{(2)}$.
In spherical symmetry, $K_2$ is proportional to $\dot\psi^2$, so
its time derivative introduces terms containing $\ddot\psi$.
For a real exponential mode, Eq.~\eqref{eq:ce-storage-match} gives
the ratio of $|\partial_vK_2|$ to
$\int\mathrm dA_0\,\mathcal P_2$ as
$[(1-\Delta_H)(5-\Delta_H)/(6\Delta_H)]\,|s|/\kappa$.
We require this ratio to be small when neglecting the storage
derivative relative to horizon absorption. At fixed $\Delta_H>0$
it vanishes in the slow limit; a small $\Delta_H$ requires a
correspondingly slower mode. The condition $|s|\ll\kappa$ also
makes the second time derivative of the entropy subleading to
$\kappa$ times its first time derivative. Under these conditions,
we retain $P_H$ when calculating the leading growth or decay rate.

The perturbation also loses energy through outgoing radiation.
As the horizon scalar value $\Phi$ changes, the scalar charge $Q(\Phi)$
changes and produces scalar waves propagating to infinity.
Using the exterior static time coordinate $t_{\rm ext}$, define the
tortoise coordinate $r_*$ by
\[
 \frac{\mathrm dr_*}{\mathrm dr}
 =\frac{e^{-\delta}}{N},
 \qquad
 v=t_{\rm ext}+r_*.
\]
With no incoming radiation, the far-zone scalar field has the outgoing
form
\[
 \phi(t_{\rm ext},r)
 =\phi_{\rm ext}
 +\frac{Q(t_{\rm ext}-r_*)}{r}
 +\cdots.
\]
At leading slow order, the horizon amplitude and far-zone scalar charge
are related by the same infinitesimal variation along the static
family, with $Q=Q(\Phi)$ and a common time normalization.
The propagation delay is represented by the retarded time
$t_{\rm ext}-r_*$.

The normalization of the scalar kinetic term in
Eq.~\eqref{eq:m-action} gives the radiated power at infinity:
\begin{equation}
 \begin{aligned}
 P_\infty
 &=\frac{1}{G}
 \left(\frac{\mathrm dQ}{\mathrm dt}\right)^2\\
 &=\frac{Q_{,\Phi}^2}{G}
 \left(\frac{\mathrm d\Phi}{\mathrm dt}\right)^2.
 \end{aligned}
 \label{eq:d-radiation}
\end{equation}
Here we used
$\mathrm dQ/\mathrm dt=Q_{,\Phi}\,\mathrm d\Phi/\mathrm dt$,
with time derivatives normalized as for the amplitude $\Phi(t)$.

The total power lost through the two channels at leading order is
\begin{equation}
 P_H+P_\infty
 =
 \frac{r_H^2\Delta_H+Q_{,\Phi}^2}{G}
 \left(\frac{\mathrm d\Phi}{\mathrm dt}\right)^2.
 \label{eq:d-total-power}
\end{equation}
The terms $r_H^2\Delta_H$ and $Q_{,\Phi}^2$ correspond to black-hole
absorption and outgoing radiation, respectively. On the regular branch
with $\Delta_H>0$, both are nonnegative and describe perturbation
energy flowing into the black hole and out to infinity.

\subsection{Slow evolution and the growth or decay rate}
\label{sec:slow-rate}

We consider a small-amplitude spherical slow mode satisfying
$|s|/\kappa\ll1$ and $|s|r_H\ll1$. In this approximation, we
neglect changes in kinetic energy and the storage terms discussed
above, and describe the mode's potential-energy change through
$E_{\rm eff}$, constructed at fixed background Wald entropy.
This static family determines the effective force. During the actual
evolution, horizon absorption produces a quadratic entropy change,
as described by Eq.~\eqref{eq:e-differential}, while outgoing scalar
radiation carries energy to infinity. The potential-energy balance
at leading slow order is therefore
\[
 \frac{\mathrm dE_{\rm eff}}{\mathrm dt}
 =-P_H-P_\infty.
\]
Substituting Eqs.~\eqref{eq:d-force} and~\eqref{eq:d-total-power} gives
\[
 \begin{aligned}
 &\frac{Q_{,\Phi}}{G}
 [\phi_\infty(\Phi)-\phi_{\rm ext}]
 \frac{\mathrm d\Phi}{\mathrm dt}\\
 &\qquad
 =-\frac{r_H^2\Delta_H+Q_{,\Phi}^2}{G}
 \left(\frac{\mathrm d\Phi}{\mathrm dt}\right)^2.
 \end{aligned}
\]
For an evolving perturbation, canceling the common factor
$\mathrm d\Phi/\mathrm dt$ and $G$ gives the slow evolution equation
\begin{equation}
 \left(r_H^2\Delta_H+Q_{,\Phi}^2\right)
 \frac{\mathrm d\Phi}{\mathrm dt}
 =-Q_{,\Phi}[\phi_\infty(\Phi)-\phi_{\rm ext}].
 \label{eq:d-slow}
\end{equation}
At this order, $r_H$, $\Delta_H$, and $Q_{,\Phi}$ take their values
on the chosen background. The source difference retains the notation
of the static family and is expanded to first order below.
Equation~\eqref{eq:d-slow} extends continuously to equilibrium. We linearize it
about an equilibrium $\Phi_0$ satisfying
$\phi_\infty(\Phi_0)=\phi_{\rm ext}$ to obtain the leading growth
or decay rate of a small perturbation. Setting
\[
 \Phi(t)-\Phi_0\propto e^{st}
\]
in Eq.~\eqref{eq:d-slow} gives
\begin{equation}
 s=-\frac{Q_{,\Phi}\phi_{\infty,\Phi}}
 {r_H^2\Delta_H+Q_{,\Phi}^2},
 \label{eq:d-rate}
\end{equation}
where all coefficients are evaluated at $\Phi_0$.
On the regular branch with $\Delta_H>0$, the denominator is positive,
so $s$ has the opposite sign to $Q_{,\Phi}\phi_{\infty,\Phi}$.
By Eq.~\eqref{eq:d-potential-curvature}, positive potential curvature
gives decay and negative curvature gives growth.

At the simple turning point, $\phi_{\infty,\Phi}=0$ and the leading
rate is $s=0$. To compare the evolution on either side, we select
nearby static black holes on both sides of the turning point and,
for each black hole, hold the external scalar source at its equilibrium
asymptotic value. Since $\phi_{\infty,\Phi\Phi}(\Phi_c)\ne0$,
$Q_{,\Phi}(\Phi_c)\ne0$, and the relevant coefficients remain regular,
the leading rate of this mode has opposite signs on the two sides,
corresponding to growth and decay. As the turning point is approached,
$|s|$ decreases and the perturbation evolves more slowly.
Equation~\eqref{eq:d-rate} gives the evolution time scale set jointly
by horizon absorption and outgoing scalar radiation.

\section{Discussion and conclusions}
\label{sec:discussion}

The membrane stress and scalar response give a coupled description of
energy exchange for perturbations of regular, nonextremal, static,
spherically symmetric EsGB black holes. Their linear energy and momentum
balances retain the structure of the Einstein-gravity membrane and
reproduce the corresponding horizon projections of the linearized metric
equations.
Scalar--curvature coupling modifies the response coefficients and links
scalar evolution to geometric expansion.

A changing horizon scalar induces expansion, so scalar work and expansion
work contribute at the same perturbative order. The horizon constraint
relates their rates, and their combined work yields the positive quadratic
power in Eq.~\eqref{eq:e-power}, with shear weight $\mathcal A$ and
scalar weight $\Delta_H$. The shear weight appears in the stress response,
whereas the scalar absorption weight emerges from the joint scalar and
expansion work. The negative bulk viscosity is compatible with positive
scalar absorption on the regular branch because the expansion rate is
constrained by the scalar evolution.

The entropy balance gives this absorbed power a thermodynamic
interpretation. For the intrinsic extension of Wald entropy specified in
Sec.~\ref{sec:entropy}, Eq.~\eqref{eq:e-law} equates the integrated
absorption between stationary cuts to the horizon temperature times the
second-order Wald-entropy change. Between nonstationary cuts, the balance
also requires the entropy-rate and temporal-storage endpoint differences.
These endpoint data, together with the integrated absorption, determine
the entropy change between general cuts.

For a single small-amplitude spherical mode near a simple turning point,
the scalar absorption coefficient determines the horizon friction for
the scalar amplitude evaluated on the evolving event horizon. In the
slow regime with no incoming radiation, combining this friction with
the effective force from a static family at fixed Wald entropy and with
outgoing monopole radiation gives the leading rate in
Eq.~\eqref{eq:d-rate}. The potential curvature fixes the linear effective
force: positive curvature gives decay and negative curvature gives growth.
For a given curvature, the sum of the positive horizon and radiative loss
coefficients sets the time scale. As the static source slope
vanishes at the turning point, the rate tends to zero and the evolution
slows. Static-family data and the independently calculated absorption
and radiation coefficients thus determine the leading rate without
fitting a dynamical damping coefficient.

Independent radial perturbation calculations and spherical evolutions
could test the predicted rate and corrections to the leading slow limit.
Combining the angular membrane responses with the exterior perturbation
equations would allow the coupled geometric and scalar contributions to
absorption and tidal response to be studied for nonspherical disturbances.
Rotating backgrounds provide a further setting in which to examine how
the horizon geometry and constraints modify these coupled responses.

\bmhead*{Acknowledgements}
This work was supported by the National Science and Technology Major Project of China (No.~2024ZD1100601), the National Key R\&D Program of China (Grant No.~2021YFC2203002), and the National Natural Science Foundation of China (Grant Nos.~12473075 and 12173071).

The authors acknowledge the assistance of OpenAI's GPT-5.5, GPT-5.6, GPT-6, and GPT-6.1 with scientific discussions, manuscript organization, and language editing. The authors take full responsibility for the scientific content and conclusions of this work.

\begin{appendices}
\numberwithin{equation}{section}
\renewcommand{\theequation}{\thesection\arabic{equation}}
\makeatletter
\let\epjcsectionfont\sectionfont
\def\sectionfont{\epjcsectionfont\def\@hangfrom##1{##1}}
\makeatother
\section{Boundary variations and horizon responses}
\label{app:boundary}

\subsection{Boundary momenta and sign conventions}

The sign of a membrane response depends on the direction of the boundary normal,
and its horizon value depends on which observer projection is taken before the
redshift limit. To make these choices explicit, we use the timelike canonical
momenta of Ref.~\cite{JulieBerti2020}, with the normal and extrinsic-curvature
conventions of Sec.~\ref{sec:model}. For the inverse metric, the
corresponding convention is
$K^{ab}=-(2\mathcal N_\perp)^{-1}\partial_w h^{ab}$ in a Gaussian chart
$ds^2=\mathcal N_\perp^2dw^2+h_{ab}dx^adx^b$, with
$n=\mathcal N_\perp^{-1}\partial_w$. Here $\mathcal N_\perp$ is the
radial lapse; it is distinct from the time lapse $\mathcal N$ used
in Sec.~\ref{sec:entropy} and Appendix~\ref{app:entropy}.
The minus sign comes from differentiating the inverse metric, not
from reversing the normal.
The on-shell coefficients are
$t_{ab}=-\pi_{ab}/(8\pi G)$ and $j_\phi=\pi_\phi/(16\pi G)$.

Define the generalized delta by
$\delta^{a_1\cdots a_p}_{b_1\cdots b_p}
=\det(\delta^{a_i}_{b_j})$. In three boundary dimensions,
\begin{align}
 \pi^a{}_b={}&-\delta^{ac}_{bd}K^d{}_c\nonumber\\
 &+4\alpha\delta^{acd}_{bij}K^i{}_cD^jD_df\nonumber\\
 &-\alpha(n^\mu\nabla_\mu f)\delta^{acd}_{bij}
          (\widehat R^{ij}{}_{cd}-2K^i{}_cK^j{}_d),
 \label{eq:b-metric}\\
 \pi_\phi={}&-4n^\mu\nabla_\mu\phi\nonumber\\
 &+2\alpha f_{,\phi}\delta^{abc}_{ijk}K^i{}_a
          (\widehat R^{jk}{}_{bc}-\tfrac23K^j{}_bK^k{}_c).
 \label{eq:b-scalar}
\end{align}

The additional rank-four delta in the general-dimensional metric
momentum vanishes in this dimension. This is why the stress depends on
derivatives of $f$, although the scalar response retains $f_{,\phi}$
times the boundary curvature completion.

For $H_{ab}=D_aD_bf$ and $H=h^{ab}H_{ab}$, the useful contractions are

\begin{align}
 &\delta^{acd}_{bij}K^i{}_cH^j{}_d\nonumber\\*
 &\quad=h^a{}_b(KH-K_{cd}H^{cd})-K^a{}_bH
             -KH^a{}_b\nonumber\\*
 &\qquad+K^a{}_cH^c{}_b+H^a{}_cK^c{}_b,\nonumber\\*
 &\delta^{acd}_{bij}\widehat R^{ij}{}_{cd}
             =-4\widehat G^a{}_b,\nonumber\\*
 &\delta^{acd}_{bij}K^i{}_cK^j{}_d
             =2(\operatorname{adj}K)^a{}_b,\nonumber\\*
 &(\operatorname{adj}K)^a{}_b=K^a{}_cK^c{}_b-KK^a{}_b\nonumber\\*
 &\hspace{24mm}+\tfrac12(K^2-K_{cd}K^{cd})\delta^a_b.
 \label{eq:b-identities}
\end{align}

The adjugate expression is polynomial and does not require invertibility
of $K^a{}_b$. For the scalar momentum,
$\delta^{abc}_{ijk}K^i{}_a\widehat R^{jk}{}_{bc}
=-4\widehat G^{ab}K_{ab}$ and
$\delta^{abc}_{ijk}K^i{}_aK^j{}_bK^k{}_c=-3J$.
Thus Eqs.~\eqref{eq:b-metric}--\eqref{eq:b-scalar} reduce to
Eqs.~\eqref{eq:m-stress}--\eqref{eq:m-scalar}.
Here $nf=n^\mu\nabla_\mu f$. Under $n\mapsto-n$, both responses
reverse sign: $K_{ab}$ and $nf$
are odd, the adjugate is even, and the Hessian is intrinsic.
The exterior Dirichlet action at its inner boundary consequently has
variation opposite to Eq.~\eqref{eq:m-variation}; the membrane response
cancels it. Variations are taken away from boundary joints.

\subsection{Observer variations and near-horizon geometry}

The observer is defined by the perturbed membrane geometry. With the
lapse and shift of Appendix~\ref{app:entropy}, its coordinate
expression is $u=\mathcal N^{-1}(\partial_v-w^A\partial_A)$.
Near the horizon $w^A=\varrho\beta^A+O(\varrho^2)$.
The background shift vanishes, so at fixed regulator
$\delta u^A=-\varrho\,\delta\beta^A/\varepsilon+O(\varepsilon^3)$
is $O(\varepsilon)$. For fixed cut coordinate vectors,
\begin{align}
 \delta(t_{ab}u^au^b)
   &=u_0^au_0^b\delta t_{ab}
                          +2t^{(0)}_{ab}u_0^a\delta u^b,\nonumber\\
 \delta(-t_{uA})
   &=-u_0^a\delta t_{aA}-t^{(0)}_{aA}\delta u^a .
 \label{eq:b-observer-variation}
\end{align}
In particular, $t^{(0)}_{AB}=p_H^{(0)}q^{(0)}_{AB}/\varepsilon
+O(\varepsilon)$. Its product with $\delta u^B$ can contribute
at order one to the momentum limit. The observer projections
in the following formulas include these variations.

The first-order horizon limits can be extracted before expanding the
individual stress components. In the observer frame of
Sec.~\ref{sec:horizon-limit}, the leading geometric terms are
\begin{equation}
\begin{aligned}
 K_{uu}^{(0)}&=-\kappa/\varepsilon+O(\varepsilon),\\
 K_{AB}^{(0)}&=\varepsilon c_Hq^{(0)}_{AB}+O(\varepsilon^3),\\
 \delta K_{AB}&=B_{AB}/\varepsilon+O(\varepsilon),\\
 \delta K_{uA}&=m_A+O(\varepsilon^2),\\
 nf&=\varepsilon d_H+\lambda\dot F/\varepsilon
              +O(\varepsilon^3)\\
   &\quad+\lambda O(\varepsilon)+O(\lambda^2),\\
 \delta H_{uu}&=\ddot F/\varepsilon^2+O(1),\\
 \delta H_{uA}&=\mathcal D_A\dot F/\varepsilon+O(\varepsilon),\\
 \delta H_{AB}&=\mathcal D_A\mathcal D_BF+O(\varepsilon^2),\\
 \widehat G_{uu,0}&=\mathcal R_0/2,\\
 \delta\widehat G_{uA}&=V_A/\varepsilon+O(\varepsilon),\\
 \delta\widehat G_{AB}
   &=\frac{\dot B_{AB}-q^{(0)}_{AB}\dot\theta}{\varepsilon^2}+O(1).
\end{aligned}
 \label{eq:b-scaling}
\end{equation}
The background intrinsic Hessian is zero because $f$ is constant on
each static membrane. Define, only for these contractions,
$W^a{}_b=\widehat G^a{}_b+(\operatorname{adj}K)^a{}_b$.
The combination in the metric response obeys
\begin{align}
 \varepsilon\delta[(nf)W^u{}_u]
   &=-\tfrac12\mathcal R_0\dot F+O(\varepsilon^2),\nonumber\\
 \delta[(nf)W^u{}_A]&=-d_HV_A+O(\varepsilon^2),\nonumber\\
 &\varepsilon\delta[(nf)W^A{}_B]\nonumber\\
 &\quad=d_H\Bigl[\dot B^A{}_B-\delta^A_B\dot\theta\nonumber\\
 &\hspace{18mm}+\kappa(\theta\delta^A_B-B^A{}_B)\Bigr]\nonumber\\
 &\qquad+\kappa c_H\dot F\delta^A_B+O(\varepsilon^2).
 \label{eq:b-normal-product}
\end{align}

These expressions show why varying the full product before taking the horizon limit is
essential. The background transverse gradient contributes the geometric deformation
terms, whereas $\delta(nf)$ produces the terms proportional to $\dot F$.
For the scalar contraction the cubic and curvature pieces give
\begin{equation}
 \varepsilon\delta(J-2\widehat G^{ab}K_{ab})
   =\kappa\delta\mathcal R
       +2c_H(\dot\theta-\kappa\theta)+O(\varepsilon^2).
 \label{eq:b-scalar-contraction}
\end{equation}
Together with $\varepsilon\delta(n^\mu\nabla_\mu\phi)\to\dot\psi$ and
$\delta f_{,\phi}=f_{,\phi\phi,H}\psi$, this gives the scalar line of
Eq.~\eqref{eq:h-response}. Equations~\eqref{eq:b-identities},
\eqref{eq:b-scaling}, and \eqref{eq:b-normal-product} give its stress
lines directly. The quadratic entropy calculation in
Appendix~\ref{app:entropy} uses the complete regulated energy balance,
with perturbation coefficients extracted before taking the horizon limit.

\subsection{Background horizon data}
\label{app:horizon-data}

For the static metric in Eq.~\eqref{eq:m-background}, expand
\begin{equation}
 \begin{aligned}
 N&=N'_H(r-\rh)+O((r-\rh)^2),\qquad N'_H>0,\\
 \delta&=\delta_H+\delta'_H(r-\rh)+O((r-\rh)^2),\\
 \phi&=\phi_H+\phi'_H(r-\rh)+O((r-\rh)^2).
 \end{aligned}
 \label{eq:m-expansion}
\end{equation}
The horizon scalar equation uses
$\Box\phi|_H=N'_H\phi'_H$. Together with the metric equations,
which give $\GB_H=12(N'_H)^2/\rh^2$, it yields
Eq.~\eqref{eq:m-regularity}. Eliminating $N'_H$ gives
\begin{equation}
 2\alpha f_{,\phi,H}\rh^2(\phi'_H)^2
 +\rh^3\phi'_H+3\alpha f_{,\phi,H}=0.
 \label{eq:m-quadratic}
\end{equation}
The roots are
$\phi'_H=\rh(-1\pm\Delta_H)/(4\alpha f_{,\phi,H})$ when
$\alpha f_{,\phi,H}\ne0$. The plus root has a smooth
vanishing-coupling limit and gives Eq.~\eqref{eq:m-horizon}.

\subsection{Scalar variation and finite membrane balances}
\label{app:finite-balances}

With the metric held fixed, direct variation of the scalar gives
\begin{equation}
 \begin{aligned}
 \delta_\phi S_D\big|_\Sigma
 ={}&\frac1{16\pi G}\int_\Sigma \mathrm d^3x\sqrt{-h}
       \Bigl[-4n^\mu\nabla_\mu\phi\\
 &+4\alpha f_{,\phi}(J-2\widehat G^{ab}K_{ab})\Bigr]\delta\phi.
 \end{aligned}
 \label{eq:m-scalar-variation}
\end{equation}

This fixes the normalization and sign of $j_\phi$ in
Eq.~\eqref{eq:m-scalar} for the normal directed out of the omitted
region. For a tangential displacement $\xi^a$ vanishing at the ends
of $\Sigma$, $\delta h^{ab}=-2D^{(a}\xi^{b)}$ and
$\delta\phi=\xi^aD_a\phi$. Inserting these variations in
Eq.~\eqref{eq:m-variation} and integrating by parts gives
\begin{equation*}
 \delta_\xi S_{\rm mem}=\int_\Sigma \mathrm d^3x\sqrt{-h}\,
 [-D_at^a{}_b+j_\phi D_b\phi]\xi^b=0,
\end{equation*}
which establishes Eq.~\eqref{eq:m-ward}.

The observers used in Eq.~\eqref{eq:m-observer-stress} are
orthogonal to the cuts, so their vorticity vanishes and
\begin{equation}
 \begin{aligned}
 a_b&=u^aD_au_b,\qquad\Theta=D_au^a,\\
 D_au_b&=-u_aa_b+\tfrac12\Theta\gamma_{ab}+\varsigma_{ab}.
 \end{aligned}
 \label{eq:m-observer-kinematics}
\end{equation}

Using $u^bt^a{}_b=-\rho u^a-\mathcal Q^a$ gives
\begin{equation}
 \begin{aligned}
 u^bD_at^a{}_b={}&-u(\rho)-(\rho+p)\Theta\\
 &-D_a\mathcal Q^a-a_a\mathcal Q^a-\Pi^{ab}\varsigma_{ab}.
 \end{aligned}
 \label{eq:m-energy-projection}
\end{equation}

Since $D_a\mathcal Q^a=\mathcal D_a\mathcal Q^a+a_a\mathcal Q^a$,
the energy projection of the Ward identity is
Eq.~\eqref{eq:m-finite-energy}. Its spatial projection is
\begin{align}
 &\gamma_b{}^cu^aD_a\mathcal Q_c
 +(\rho+p)a_b+\mathcal D_bp+\mathcal D_a\Pi^a{}_b\nonumber\\
 &\quad+\Pi_{ba}a^a+\tfrac32\Theta\mathcal Q_b
       +\varsigma_{ba}\mathcal Q^a=j_\phi\mathcal D_b\phi.
 \label{eq:m-finite-momentum}
\end{align}

The factor $3/2$ follows from the two spatial dimensions of the
observer cuts. These are projections of the common boundary
identity and contain the scalar work and force.

\subsection{Derivation of the linear responses and horizon balances}
\label{app:linear-details}

The contractions in Sec.~\ref{sec:horizon-limit} give the spatial
stress variation before its trace is extracted:
\begin{align}
 8\pi G\,\delta s_{HAB}={}&\kappa\chi_{AB}
       +\mathcal A(\theta q^{(0)}_{AB}-B_{AB})\nonumber\\
 &-4\alpha d_H(\dot B_{AB}-q^{(0)}_{AB}\dot\theta)\nonumber\\
 &+4\alpha\kappa(\mathcal D_A\mathcal D_BF
                    -q^{(0)}_{AB}\mathcal D^2F)\nonumber\\
 &+4\alpha c_H(\ddot F-\kappa\dot F)q^{(0)}_{AB}.
 \label{eq:h-derived-spatial}
\end{align}

The trace must include the perturbed inverse metric,
\begin{equation}
 \delta p_H=\tfrac12q_{(0)}^{AB}\delta s_{HAB}
                         -\tfrac12p_H^{(0)}\chi,
 \label{eq:h-pressure-extraction}
\end{equation}
so the trace of $\kappa\chi_{AB}$ cancels the metric-variation
term. The complete covariant stress variation is reconstructed as
\begin{equation}
 \delta s_{HAB}=p_H^{(0)}\chi_{AB}
 +\delta p_Hq^{(0)}_{AB}+\delta\tau_{AB}.
 \label{eq:h-spatial}
\end{equation}
The observer variations needed for the momentum are included in
Eq.~\eqref{eq:b-observer-variation}.

For the scalar response, $J=-2\det(K^a{}_b)$ in three boundary
dimensions. The regulated background $\varepsilon J^{(0)}$ vanishes,
whereas
$\varepsilon(J-2\widehat G^{ab}K_{ab})^{(0)}
=\kappa\mathcal R_0+O(\varepsilon^2)$.
The cubic extrinsic-curvature and intrinsic-curvature contributions to
Eq.~\eqref{eq:b-scalar-contraction} are separately
\begin{align}
 \lim_{\varepsilon\to0}\varepsilon\delta J
    &=-2\kappa c_H\theta,\nonumber\\*
 \lim_{\varepsilon\to0}
       \varepsilon\delta(-2\widehat G^{ab}K_{ab})
    &=\kappa\delta\mathcal R+2c_H\dot\theta .
 \label{eq:h-scalar-pieces}
\end{align}
Their sum gives Eq.~\eqref{eq:b-scalar-contraction}, completing the
curvature contribution to the scalar response in Eq.~\eqref{eq:h-response}.

Substitution into the linear membrane energy equation gives
\begin{equation}
 -\dot\theta-2\alpha\mathcal R_0\ddot F+\kappa\theta
                  =-2\alpha\kappa\mathcal R_0\dot F,
 \label{eq:h-energy-substitution}
\end{equation}
which factors into Eq.~\eqref{eq:h-null}. For the momentum equation,
\begin{align}
 &8\pi G(\mathcal D_A\delta p_H
                    +\mathcal D^B\delta\tau_{AB})\nonumber\\
 &\quad=-\mathcal A V_A-4\alpha d_H\dot V_A
                    +4\alpha c_H\mathcal D_A\ddot F\nonumber\\
 &\qquad-4\alpha\kappa c_H\mathcal D_A\dot F
                    +2\alpha\kappa\mathcal R_0\mathcal D_AF.
 \label{eq:h-stress-divergence}
\end{align}

The $\dot V_A$ and $\ddot F$ terms cancel those in
$\partial_v\delta\mathcal P_A$. The sphere identity
\begin{equation}
 \mathcal D^B(\mathcal D_A\mathcal D_BF
    -q^{(0)}_{AB}\mathcal D^2F)
 =\tfrac12\mathcal R_0\mathcal D_AF,
 \label{eq:h-angular-identity}
\end{equation}
accounts for the curvature-dependent scalar force, leaving
Eq.~\eqref{eq:h-codazzi}.

For the direct bulk check, the null projection uses
$P_{v\varrho v\varrho}^{(0)}=-\mathcal R_0/2$ and
$\delta(\nabla^\varrho\nabla^\varrho f)=\ddot F-\kappa\dot F$.
The angular projection additionally uses
\begin{align}
 \delta P_{v\varrho vA}&=V_A,
 &P^{(0)}_{vBA\varrho}&=\kappa c_Hq^{(0)}_{AB},\nonumber\\
 (\nabla^\varrho\nabla^vf)_0&=\kappa d_H,
 &\delta(\nabla^B\nabla^\varrho f)&=\mathcal D^B\dot F.
 \label{eq:h-bulk-contractions}
\end{align}
The minimally coupled scalar stress has no linear contribution to
either projection on the static horizon. The resulting bulk equations
are Eq.~\eqref{eq:h-bulk}.

\section{Membrane energy and entropy balance}
\label{app:entropy}

The finite-cut entropy balance requires the integrated membrane energy to have a
well-defined perturbative horizon limit, even though individual terms may diverge
on a stretched surface. We establish that limit through quadratic order and
compare the resulting scalar power and storage with the spherical bulk null
equation. Throughout, perturbation coefficients are extracted at nonzero membrane
lapse before the horizon limit is taken.

\subsection{Integrated membrane energy and entropy rate}

Write the induced metric as
$h_{ab}\mathrm dx^a\mathrm dx^b=-\mathcal N^2\mathrm dv^2+
q_{AB}(\mathrm dx^A+w^A\mathrm dv)(\mathrm dx^B+w^B\mathrm dv)$,
so that $u=\mathcal N^{-1}(\partial_v-w^A\partial_A)$.
The lapse and shift are those of Sec.~\ref{sec:entropy}, with
$\mathcal N^2=\varepsilon^2+O(\varrho^2)$ and
$w^A=\varrho q^{AB}\beta_B+O(\varrho^2)$. Until the perturbation
coefficients are extracted, $q_{AB}$ is the full cut metric,
$\mathcal D_A$ is its compatible connection, and $\mathcal R$ is its
Ricci scalar. Spatial indices are raised with $q^{AB}$, and a dot
denotes $\partial_v$ at fixed regulator.
Let $\mathcal Q^A$ denote the spatial energy flux in this appendix,
and let $a_A=\mathcal D_A\ln\mathcal N$ be the observer acceleration.
Their contribution to the integrated energy equation vanishes exactly:
\begin{equation}
 \begin{aligned}
 &\int\mathrm d^2x\sqrt q\,\mathcal N^2
       (\mathcal D_A\mathcal Q^A+2a_A\mathcal Q^A)\\
 &\quad=\int\mathrm d^2x\sqrt q\,
       \mathcal D_A(\mathcal N^2\mathcal Q^A)=0.
 \end{aligned}
 \label{eq:ce-flux}
\end{equation}

Because this identity is evaluated before taking the regulator limit, possible poles in
the quadratic spatial flux cancel within the integrated divergence. The density terms
become
$\partial_v\mathcal E_{\rm mem}$ plus a remainder proportional to
$\mathcal N\rho(\partial_v-w^A\mathcal D_A)\ln\mathcal N$.
To determine whether this remainder survives the horizon limit,
we first examine the perturbative orders of $\mathcal N\rho$.

For the density contraction, let $L_{AB}$ be the extrinsic curvature
of a cut within the timelike membrane and
$C_{AB}=\partial_\varrho q_{AB}/2$ its transverse deformation. Then
\begin{equation}
 \begin{aligned}
 K_{AB}&=L_{AB}+\mathcal NC_{AB},\\
 D_AD_Bf&=\mathcal D_A\mathcal D_Bf-L_{AB}uf,\\
 nf&=uf+\mathcal N\partial_\varrho f,\\
 \widehat G_{uu}&=\frac{\mathcal R}{2}+\det L.
 \end{aligned}
 \label{eq:ce-geometry}
\end{equation}

Here $uf=u^aD_af$, $nf=n^\mu\nabla_\mu f$, and
$\det L=\det(L^A{}_B)$ is the two-dimensional determinant.
In the canonical density, the scalar Hessian and
normal-derivative terms contain opposite contributions proportional to
$\mathcal Nuf[(\operatorname{tr}L)(\operatorname{tr}C)
-\operatorname{tr}(LC)]$. Their sum instead contains
$nf-uf=\mathcal N\partial_\varrho f$ and vanishes after the density is
multiplied by $\mathcal N$ and the limit is taken.

Define the full shifted rates
$\overline{\mathcal B}_{AB}=\dot q_{AB}/2-
\mathcal L_wq_{AB}/2=\mathcal NL_{AB}$,
$\bar\vartheta=q^{AB}\overline{\mathcal B}_{AB}$ and
$\overline{\dot f}=\dot f-w^A\mathcal D_Af$. Here
$\mathcal L_w$ denotes the Lie derivative along the spatial shift
$w=w^A\partial_A$, so that
$(\mathcal L_wq)_{AB}=\mathcal D_Aw_B+\mathcal D_Bw_A$, with
$w_A=q_{AB}w^B$.
The remaining density is
\begin{align}
 &\mathcal N\rho=-\frac1{8\pi G}
       \Bigl[\bar\vartheta+2\alpha\mathcal R\overline{\dot f}\nonumber\\
 &\hspace{16mm}+4\alpha
       (\overline{\mathcal B}^{AB}\mathcal D_A\mathcal D_Bf
                           -\bar\vartheta\mathcal D^2f)\Bigr]\nonumber\\
 &\quad-\frac{\alpha\overline{\dot f}}{2\pi G\mathcal N^2}
       (\bar\vartheta^2-\overline{\mathcal B}_{AB}
                        \overline{\mathcal B}^{AB})+O(\mathcal N^2).
 \label{eq:ce-density}
\end{align}

Each shifted rate vanishes on the stationary background, so the pole
is $O(\lambda^3/\mathcal N^2)$ and does not enter the coefficients
through quadratic order when they are extracted at fixed nonzero lapse.
Those coefficients of $\mathcal N\rho$ are therefore bounded as
$\mathcal N\to0$. In the fixed-nonaffinity gauge, the coefficients
of $(\partial_v-w^A\mathcal D_A)\ln\mathcal N$ are
$O(\mathcal N^2)$ through the same order. Their products with the
bounded density coefficients consequently vanish, which removes
the lapse-gradient remainder in the integrated energy equation.
The shift-dependent corrections to the finite bracket also vanish
in the horizon limit. Through quadratic order we may therefore use
the unshifted horizon rates
$\mathcal B_{AB}=\dot q_{AB}/2$ and
$\vartheta=q^{AB}\mathcal B_{AB}$.

The intrinsic-curvature variation on a closed cut gives
\begin{equation}
 \begin{aligned}
 \dot S_{\rm intr}={}&\frac1{4G}\int\mathrm d^2x\sqrt q
       \Bigl[\vartheta+2\alpha\mathcal R\dot f\\
 &+4\alpha(\mathcal B^{AB}\mathcal D_A\mathcal D_Bf
                         -\vartheta\mathcal D^2f)\Bigr].
 \end{aligned}
 \label{eq:ce-entropy-rate}
\end{equation}

It follows from $\dot{\sqrt q}=\sqrt q\vartheta$ and
$\dot{\mathcal R}=-\mathcal R\vartheta+
2\mathcal D_A\mathcal D_B\mathcal B^{AB}-2\mathcal D^2\vartheta$,
followed by two integrations by parts. Comparing prefactors in
\eqref{eq:ce-density} and \eqref{eq:ce-entropy-rate} proves
\eqref{eq:e-density-storage}. This comparison includes the quadratic
area measure and the full Taylor expansion of $f$.

To identify the equilibrium work, we expand the scalar canonical
density. With $\chi=q_{(0)}^{AB}\chi_{AB}$, its first-order variation is
given below; indices and angular derivatives use the background metric
$q^{(0)}_{AB}$:
\begin{equation}
 \begin{aligned}
 &\frac{\delta(\sqrt q\,j_{\rm eq})}{\sqrt{q^{(0)}}}\\
 &\quad=\frac{\alpha\kappa}{4\pi G}\Bigl[f_{,\phi,H}
       (\mathcal D_A\mathcal D_B\chi^{AB}-\mathcal D^2\chi)\\
 &\hspace{40mm}+f_{,\phi\phi,H}\mathcal R_0\psi\Bigr].
 \end{aligned}
 \label{eq:ce-scalar-density}
\end{equation}

The trace part of the curvature variation cancels the area variation;
the angular derivatives remain. Keeping this canonical density and
the spatial Hessian work together gives exactly
$\partial_vI_{\rm eq}=T_0\dot S_{\rm intr}$ in the integrated equation.
The residual work then uses only the first-order responses. In
\eqref{eq:e-work}, the mixed derivative terms combine as
$\theta\ddot\psi+\dot\theta\dot\psi=
\partial_v(\theta\dot\psi)$, fixing the mixed storage coefficient
in \eqref{eq:e-storage}.

\subsection{Entropy balance in spherical symmetry}

For spherical evolution use
\begin{equation}
 \begin{aligned}
 \mathrm ds^2={}&-\mathsf F(v,\varrho)\mathrm dv^2
     +2\mathrm dv\mathrm d\varrho+r(v,\varrho)^2\mathrm d\Omega_2^2,\\
 \mathsf F(v,0)&=0,\qquad
       \partial_\varrho\mathsf F(v,0)=2\kappa(v).
 \end{aligned}
 \label{eq:ce-spherical-metric}
\end{equation}

All quantities below are evaluated at $\varrho=0$, and
$r_{,\varrho},f_{,\varrho}$ are full evolving transverse derivatives.
The $vv$ metric equation reduces to
\begin{align}
 &[r-4\alpha(\dot r f_{,\varrho}+r_{,\varrho}\dot f)]
                  (\ddot r-\kappa\dot r)\nonumber\\
 &\qquad+2\alpha(1-2\dot r r_{,\varrho})
                  (\ddot f-\kappa\dot f)+r^2\dot\phi^2=0.
 \label{eq:ce-null}
\end{align}
For this metric $S_{\rm intr}=\pi W/G$, where
$W=r^2+4\alpha f$. Rearranging \eqref{eq:ce-null} yields
\begin{align}
 &(\partial_v-\kappa)\dot W
       =-2r^2\dot\phi^2+2\dot r^2\nonumber\\
 &\quad+8\alpha(\ddot r-\kappa\dot r)
                  (\dot r f_{,\varrho}+r_{,\varrho}\dot f)\nonumber\\
 &\quad+8\alpha\dot r r_{,\varrho}(\ddot f-\kappa\dot f).
 \label{eq:ce-W}
\end{align}

Expand $r=\rh+\lambda r^{(1)}+\lambda^2r^{(2)}$, and similarly
$\phi=\phi_H+\lambda\psi+\lambda^2\psi^{(2)}$ and
$W=W^{(0)}+\lambda W^{(1)}+\lambda^2W^{(2)}$.
Write the nonaffinity as $\kappa(v)=\kappa+\lambda\kappa^{(1)}+O(\lambda^2)$,
where $\kappa$ is its background value.
The first-order equation is
$(\partial_v-\kappa)\dot W^{(1)}=0$.
For perturbations that approach future stationarity, the future
condition excludes its growing homogeneous solution. The slow modes
of Sec.~\ref{sec:dynamical-input} also satisfy $\dot W^{(1)}=0$
directly, since their exponential rate obeys $|s|\ll\kappa$.
In either case,
$\dot W^{(1)}=2\rh\dot r^{(1)}+4\alpha f_{,\phi,H}\dot\psi=0$.
At second order the entropy combination is
\begin{equation}
 \begin{aligned}
 W^{(2)}={}&\bigl(r^{(1)}\bigr)^2+2\rh r^{(2)}\\
 &+4\alpha f_{,\phi,H}\psi^{(2)}
               +2\alpha f_{,\phi\phi,H}\psi^2.
 \end{aligned}
 \label{eq:ce-W2}
\end{equation}

Only background transverse values
$r_{,\varrho}=\rh c_H$ and $f_{,\varrho}=d_H$ enter the quadratic
source. Substitution of
$\dot r^{(1)}=-2\alpha f_{,\phi,H}\dot\psi/\rh$ gives
\begin{equation}
 \begin{aligned}
 (\partial_v-\kappa)\dot W^{(2)}
 ={}&-2\rh^2\Delta_H\dot\psi^2\\
 &+\frac{32\alpha^2 f_{,\phi,H}^{\,2}}{\rh^2}
       (\alpha d_H-\rh^2c_H)\dot\psi\ddot\psi.
 \end{aligned}
 \label{eq:ce-quadratic}
\end{equation}

Here we have used
$\kappa(\alpha d_H-\rh^2c_H)=(1-3\mathcal A)/4$
and the regularity relation $\mathcal A=2/(1+\Delta_H)$ to simplify
the coefficient of $\dot\psi^2$ according to
\[
 1-\frac{12\alpha^2 f_{,\phi,H}^{\,2}\mathcal A}{\rh^4}=\Delta_H.
\]
A varying nonaffinity contributes
$-\kappa^{(1)}\dot W^{(1)}=0$ at this order.
In the spherical limit, \eqref{eq:e-storage} integrates to
\begin{equation}
 K_2=\frac{8\alpha^2 f_{,\phi,H}^{\,2}}{G\rh^2}
                          (\alpha d_H-\rh^2c_H)\dot\psi^2.
 \label{eq:ce-storage-match}
\end{equation}
Multiplication of \eqref{eq:ce-quadratic} by $\pi/G$ therefore recovers
the scalar absorption term and the temporal-storage derivative in
Eq.~\eqref{eq:e-differential}. With stationary endpoints it
gives $T_0\Delta S_W^{(2)}=(\rh^2\Delta_H/G)
\int\mathrm dv\,\dot\psi^2$, the scalar part of \eqref{eq:e-law}.

\makeatletter
\let\sectionfont\epjcsectionfont
\makeatother
\end{appendices}

\end{document}